\documentclass[12pt]{article}
\usepackage{ifpdf}
\ifpdf
\usepackage{graphicx,color}
\usepackage{hyperref}
\else
\usepackage[dvipdfmx]{graphicx,color}
\usepackage[dvipdfmx]{hyperref}
\fi
\usepackage{amssymb,amsfonts,amsmath,cancel,cite,multirow}
\usepackage[capitalise]{cleveref}

\newcommand{\pdiff}[3]{
\if 1#1   \frac{\partial #2 }{\partial #3 }
\else  \frac{\partial^#1#2 }{\partial #3^#1 } \fi}
\newcommand{\diff}[3]{
\if 1#1  \frac{\mathrm{d} #2 }{\mathrm{d} #3 }
\else  \frac{\mathrm{d}^{#1} #2 }{\mathrm{d}#3^{#1} } \fi
}

\newcommand{\Tr}[0]{\mathrm{Tr}}

\newcommand{\ave}[1]{\left \langle #1 \right \rangle}
\newcommand{\evalue}[3]{\left \langle #1 \right| #3 \left | #2 \right \rangle}
\newcommand{\diag}{\text{diag}}
\newcommand{\sign}{\text{sign}}

\newcommand{\Lcal}[0]{\mathcal{L}}
\newcommand{\Ocal}[0]{\mathcal{O}}
\newcommand{\MPl}[0]{M_{\text{Pl}}}
\newcommand{\MGUT}[0]{M_{\mathrm{GUT}}}
\newcommand{\MSUSY}[0]{M_{\mathrm{SUSY}}}
\newcommand{\UMNS}[0]{U_{\mathrm{PMNS}}}

\newcommand{\MSbar}[0]{$ \ovl{\text{MS}}~$}
\newcommand{\DRbar}[0]{$ \ovl{\text{DR}}~$}
\newcommand{\FH}[0]{\texttt{FeynHiggs 2.14.2}}

\newcommand{\lam}{\lambda}
\newcommand{\vph}[0]{\varphi}
\newcommand{\ep}{\epsilon}
\newcommand{\del}{\delta}
\newcommand{\Del}{\Delta}

\newcommand{\Si}{\Sigma}
\newcommand{\ovl}{\overline}
\newcommand{\eqs}[1]{\begin{equation}\begin{split} #1 \end{split}\end{equation}}

\begin{document}

\begin{titlepage}

\begin{flushright}
	CTPU-PTC 18-21 \\
	KIAS-P18081
\end{flushright}

\vskip 1.35cm
\begin{center}

{\large
\textbf{
Revisiting Flavor and CP Violation in Supersymmetric $SU(5)$ with Right-Handed Neutrinos
}}
\vskip 1.2cm

Jason L. Evans$^a$,
Kenji Kadota$^b$,
and Takumi Kuwahara$^b$\\

\vskip 0.4cm

\textit{$^a$
School of Physics, KIAS, Seoul 130-722, Korea}\\
\textit{$^b$
Center for Theoretical Physics of the Universe, Institute for Basic Science (IBS), Daejeon 34126, Korea}

\vskip 1.5cm

\begin{abstract}
	We revisit the minimal supersymmetric $SU(5)$ grand unified theory with three right-handed neutrinos in which universality conditions for soft-supersymmetry breaking parameters are imposed at an input scale above the unification scale.
	If the Majorana masses for the neutrinos are around $10^{15}~\mathrm{GeV}$, large mixing angles and phases in the neutrino sector lead to flavor-violation and $CP$-violation in the right-handed down squark and left-handed slepton sectors.
	Since the observed Higgs boson mass and the proton decay constraints indicate sfermions have masses larger than a few TeV, flavor and $CP$ constraints are less restrictive.
	We explore the constraints on models with a universal soft-supersymmetry breaking input parameters coming from proton stability, electric dipole moments, $\mu\to e\gamma$ decays, and the Higgs mass observed at the LHC. Regions compatible with all constraints can be found if non-zero $A$-terms are taken.

\end{abstract}

\end{center}
\end{titlepage}

\section{Introduction}

The collider experiments at the Large Hadron Collider (LHC) have given stringent constraints on models beyond the Standard Model (SM).
In the minimal supersymmetric (SUSY) extension of the SM (MSSM), squarks and gluinos are severely constrained by their absence at the LHC (see Refs.~\cite{Aad:2015iea,Aaboud:2017vwy,Sirunyan:2017bsh}) and by the observed Higgs mass \cite{Aad:2015zhl}.

In many models, the soft-supersymmetry breaking parameters are assumed to be universal and real at the input scale.
This assumption makes it easy to avoid constraints from flavor-changing and $CP$-violating processes (such as meson oscillations, rare decays, electric dipole moments (EDMs), etc).
For most studies, the universality conditions on the soft-supersymmetry breaking parameters are imposed at the grand unification (GUT) scale, $M_{\text{GUT}} \sim 2 \times 10^{16}~\text{GeV}$, where the SM gauge couplings unify.
In particular, in the constrained MSSM (CMSSM) \cite{Nilles:1983ge,Kane:1993td} the universality of the scalar mass ($m_0$), gaugino mass ($M_{1/2}$), and trilinear coupling ($A_0$) are assumed at the unification scale.
In this simplified model, flavor and $CP$ violating processes arise only through the Cabbibo-Kobayashi-Maskawa (CKM) matrix and are suppressed due to the smallness of the CKM matrix elements.
However, there is no compelling reason to take the boundary scale of the soft-supersymmetry breaking parameters to be the GUT scale, and it is quite plausible that it is above the GUT scale (the so-called super-GUT scenarios, see Refs.~\cite{Ellis:2010jb,Ellis:2010ip,Dudas:2012hx,Ellis:2016tjc}) or below the GUT scale (the so-called sub-GUT scenarios, see Refs.~\cite{Ellis:2018jyl,Ellis:2015rya,Ellis:2007ac,Ellis:2006vc}).
In the case of super-GUT models, the low-scale soft-supersymmetry breaking parameters are affected by which SUSY GUT model is chosen, because of the renormalization group (RG) running between the input scale and the unification scale.

Grand unified theories based on $SU(5)$ are among the more minimal options; each generation of quarks and leptons are unified into a $\ovl{\mathbf{5}} + \mathbf{10}$ representations of $SU(5)$ \cite{Georgi:1974sy,Sakai:1981gr,Dimopoulos:1981zb}.
Although neutrinos are massless in the SM, non-zero neutrino masses and mixing angles have been well established by the neutrino oscillation experiments \cite{Gonzalez-Garcia:2014bfa,Esteban:2016qun,Capozzi:2017ipn}.
In the context of $SU(5)$ GUTs, tiny neutrino masses are realized by introducing a singlet fermion, the right-handed neutrinos, and relying on the type-I seesaw mechanism \cite{Minkowski:1977sc,Mohapatra:1979ia,GellMann:1980vs,Schechter:1980gr}.
Adding right-handed neutrinos to the MSSM introduces large flavor mixing in the neutrino sector, which in turn induces large flavor changing processes in the charged lepton sector \cite{Leontaris:1985pq,Borzumati:1986qx,Hisano:1995cp,Hisano:1995nq,Hisano:1998fj,Calibbi:2006nq}.
For super-GUT models with right-handed neutrinos, large flavor changing processes arise in the down-quark sector as well, due to the down-quark's interaction with color-triplet Higgs field and the right-handed neutrinos \cite{Baek:2000sj,Moroi:2000tk,Akama:2001em,Moroi:2000mr,Chang:2002mq,Hisano:2003bd,Ciuchini:2003rg,Hisano:2004pw,Buras:2010pm}.

In this paper, we revisit the super-GUT CMSSM scenario with three right-handed neutrinos.
The current best fit of the neutrino oscillation experiments have revealed that the neutrino sector has large mixing angles and most likely a large non-zero Dirac $CP$ phase \cite{Gonzalez-Garcia:2014bfa,Esteban:2016qun,Capozzi:2017ipn}.
Including right-handed neutrinos in super-GUT models, the resulting flavor and $CP$ violating effects give significant constraints on the soft-supersymmetry breaking input parameters.
Although maintaining proton stability is challenging in minimal $SU(5)$ SUSY GUTs with multi-TeV scale SUSY particles \cite{Goto:1998qg,Murayama:2001ur}, the lifetime of the proton can be pushed beyond current experimental bounds by properly choosing the additional Yukawa couplings and $CP$ phases \cite{Ellis:2016tjc}, even for multi-TeV soft SUSY masses.
Here, we examine the parameter space of the boundary masses, $(m_0, M_{1/2}, A_0)$, in a super-GUT CMSSM which are compatible with the Higgs mass, flavor changing, and $CP$ violating constraints.
More particularly, we examine constraints from: meson oscillations, lepton flavor violation, electric dipole moments, and proton decay.

Previous studies revealed that the sparticle mass spectrum could change in the presence of right-handed neutrinos, even if CMSSM boundary conditions are assumed at the GUT scale \cite{Moroi:1993ui,Kadota:2009vq,Kadota:2009fg}.
For right-handed neutrino Majorana masses of around $10^{15}~\mathrm{GeV}$, the neutrino Yukawa couplings are $\Ocal(1)$ and they contribute significantly to the renormalization group equations (RGEs) of the MSSM soft-supersymmetry breaking parameters.
As a result, the left-handed tau slepton can be the next-lightest supersymmetric particle (NLSP) instead of right-handed sleptons.
For super-GUT boundary conditions, the soft mass for right-handed sneutrinos and down-squarks are also affected by the large neutrino Yukawa couplings above the GUT scale.
The effect of this added running on the MSSM soft masses is not obvious and must be investigated in detail.

This paper is organized as follows:
in \cref{sec:model}, we will briefly review the minimal SUSY $SU(5)$ GUT with three right-handed neutrinos and its associated soft-supersymmetry breaking parameters.
We then discuss the constraints on flavor and $CP$ violation of this model, in particular we examine the effect of meson mixing, lepton flavor violating decays $\mu\to e\gamma$, EDMs, and proton stability.
Our results are presented in \cref{sec:result}, and finally, we summarize our work in \cref{sec:conclusion}.

\section{Model \label{sec:model}}

First, we review the minimal SUSY $SU(5)$ GUT with three right-handed neutrinos and fix our notation.

The MSSM matter fields are contained in three $\ovl{\mathbf{5}} + \mathbf{10}$ representations of $SU(5)$.
The chiral multiplets $\Phi_i$ are in the $\ovl{\mathbf{5}}$ representation.
They are composed of the right-handed down-type quark multiplets $\ovl D_i$ and the lepton doublets $L_i$, while $\Psi_i$ is in the $\mathbf{10}$ representation and is composed of the quark doublets $Q_i$, the right-handed up-type quark multiplets $\ovl U_i$, and the right-handed charged leptons $\ovl E_i$.
The right-handed neutrinos $\ovl N_i$ are singlets of $SU(5)$.
The subscript $i$ on all fields denotes the generation.
The MSSM Higgs doublets, $H_u$ and $H_d$, are incorporated into the fundamental and anti-fundamental representations of $SU(5)$, $H$ and $\ovl H$, with the color-triplet Higgs superfields, $H_C$ and $\ovl H_C$ respectively.
The $SU(5)$ gauge group is spontaneously broken by the vacuum expectation value (VEV) of the $\mathbf{24}$ Higgs multiplet $\Si$.

\subsection{Superpotential}
The superpotential for our model is given by
\eqs{
W & = \frac{f^u_{ij}}{4} \Psi_i \Psi_j H
+ \sqrt 2 f^d_{ij} \Psi_i \Phi_j \ovl H
+ f^\nu_{ij} \Phi_i \ovl N_j H
+ \frac12 (M_R)_{ij} \ovl N_i \ovl N_j
+ W_{\text{GUT}} + W_{\text{Pl}} \, ,
\label{eq:superpotential}
}
Here, $f^u$ and $f^d$ contain the MSSM Yukawa couplings, and $f^\nu$ is the neutrino Yukawa coupling.
$M_R$ is the Majorana mass matrix for the right-handed neutrinos.
The $SU(5)$ indices are suppressed in \cref{eq:superpotential}.
$W_{\text{GUT}}$ represents the superpotential couplings among the Higgs fields $H, \ovl H$, and $\Si$, while $W_{\text{Pl}}$ contains the Planck-scale suppressed operators up to dimension five.

The Higgs-sector superpotential is given by
\eqs{
W_{\text{GUT}} & = \mu_H \ovl H H
+ \lam \ovl H \Si H
+ \frac{\mu_\Si}{2} \Tr \Si^2
+ \frac{\lam_\Si}{3} \Tr \Si^3 \, .
}
The adjoint Higgs $\Si = \Si^a T^a$, where $T^a~(a = 1, \cdots, 24)$ is the generator of $SU(5)$, is assumed to have a VEV of the form; $\ave{\Si} = v \, \diag(2,2,2,-3,-3)$ with $v = \mu_\Si/\lam_\Si$.
After symmetry breaking, the SU(5) gauge bosons acquire masses $M_X = 5 \sqrt 2 g_5 v$, where $g_5$ is the $SU(5)$ gauge coupling.
$\mu_H$ must be fine-tuned to realize the doublet-triplet splitting, $\mu_H = 3 \lam v$, and then the color-triplet Higgs superfields obtain masses $M_{H_C} = 5 \lam v$.
The color-octet and weak-triplet components of $\Si$ get a mass $M_\Si = \frac52 \lam_\Si v$, while the SM singlet component has a mass $M_\Si/5$.

The higher-dimensional operators which involve the adjoint chiral superfield $\Si$ can play a significant role in the matching conditions at the GUT scale.
The following is the superpotential containing the important operators for this work,
\eqs{
W_{\text{Pl}} & = \frac{c_{ij}}{\MPl} \Psi_i^{AC} \Si_A^B \Phi_{jB} \ovl H_C
+ \frac{c^\prime_{ij}}{\MPl} \Phi_{iA} \Si_B^A \ovl N_j H^B \\
& ~~~~
+ \frac{a_{ij}}{\MPl} (\Tr \Si^2) \ovl N_i \ovl N_j
+ \frac{c}{\MPl} \Tr [\Si \mathcal{W} \mathcal{W}] \, .
\label{eq:Pl_supop}
}
Here, $\MPl = 2.4 \times 10^{18}~\mathrm{GeV}$ is the reduced Planck mass and capital letters, $A,B,C,\cdots = 1 \mathrm{-} 5$, denote the $SU(5)$ indices.
The first term is introduced in order to realize the correct matching conditions for the down-Yukawa and lepton-Yukawa couplings \cite{Ellis:1979fg,Panagiotakopoulos:1984wf,Bajc:2002pg}
\footnote{
There are other operators involving $\Si$, such as $\Psi_i^{AB} \Phi_{jA} \Si_B^C \ovl H_C $ and $\Psi_i \Psi_j \Si H$.
However, they only modify the Yukawa couplings with the color-triplet Higgs fields.
}.
The second term changes the matching conditions for $f^\nu$, and the third term gives corrections of order $\Ocal(v^2/\MPl)$ to the Majorana mass of the right-handed neutrinos.
Since we assume a right-handed Majorana mass of order $10^{15}~\mathrm{GeV}$, these corrections are negligible, which is confirmed in our numerical studies.
The last term in \cref{eq:Pl_supop} involving $\mathcal{W}$, the field-strength chiral superfield of $SU(5)$, gives a finite correction to the gauge coupling matching conditions at the GUT scale.
\footnote{
If the higher-dimensional operators $(\Tr \Si^2)^2$ and $\Tr \Si^4$ are included, the color octet's and weak triplet's masses are split by an amount $\Ocal(v^2/\MPl)$.
As long as the coefficients of these operators are not too large, they can be safely ignored.
}

Using unitary transformations in flavor space, the MSSM superfields in the mass basis, where the Yukawa matrices for up-type quarks and charged leptons are diagonalized, are given by
\eqs{
Q_i & = \left(
\begin{array}{c}
	U_i \\
	V_{ij} D_j
\end{array} \right) \, , ~~~
e^{- i \vph_{u_i}} \ovl U_i \, , ~~~
V_{ij} \ovl E_j \in \Psi_i \, , \\
L_i & = \left(
\begin{array}{c}
	e^{- i \vph_{d_i}} U_{ij} N_j \\
	E_i
\end{array} \right) \, , ~~~
\ovl D_i \,  \in \Phi_i \, ,
}
and then the unphysical degrees of freedom in the Yukawa couplings and Majorana mass matrices can be removed,
\eqs{
f^u_{ij} & = f^u_i e^{i \vph_{u_i}} \del_{ij} \, , \\
f^d_{ij} & = f^d_i V^\ast_{ij} \, , \\
f^\nu_{ij} & = f^\nu_j e^{i \vph_{d_i}} U^\ast_{ij} \, , \\
(M_R)_{ij} & = e^{i \vph_{\nu_i}} W_{ik} (M_R^D)_k e^{2 i \ovl\vph_{\nu_k}} W_{jk} e^{i \vph_{\nu_j}} \, ,
}
where $f^u_i \,, f^d_i \,, f^\nu_i$ , and $(M_R^D)_i$ are the eigenvalues of the corresponding matrices.
$V$ is identified with the CKM matrix, and the other unitary matrices, $U$ and $W$, are defined such that they contain only a single $CP$ phase.
If $(M_R)_{ij}$ is taken to be diagonal, the unitary matrix $W$ is unity and $U$ can then be identified with the Pontecorvo-Maki-Nakagawa-Sakata (PMNS) matrix.
$\vph_{f_i}~(f = u, d, \nu)$ are additional phases present above the GUT scale and are constrained as follows, $\vph_{f_1} + \vph_{f_2} + \vph_{f_3}= 0$.

The mass matrix for the left-handed neutrinos comes from the dimension-five operator generated when the right-handed neutrinos are integrated out of \cref{eq:superpotential}.
After electroweak symmetry breaking (EWSB), the operator gives a neutrino mass matrix of
\eqs{
(m_\nu)_{ij} = \frac{v_u^2}{2} (f^\nu M_R^{-1} f^{\nu T})_{ij} \,.
}
This mass matrix is diagonalized by the PMNS matrix $\UMNS$,
\eqs{
m_\nu^{\diag} \equiv \diag(m_1, m_2, m_3) = \UMNS^T m_\nu \UMNS \,,
}
where $m_i$ denote the mass eigenvalues.
$\UMNS$ is characterized by three mixing angles, $\theta_{12} \,, \theta_{23} \,,$ and $\theta_{31} \,$, a Dirac $CP$ phase $\del_{CP}$, and a diagonal matrix containing the Majorana phases.
In our study, we use the central values for the mixing angles and the Dirac $CP$ phase found in Ref.~\cite{Patrignani:2016xqp},
\eqs{
\sin^2\theta_{12} = 0.297 \, , ~~~
\sin^2\theta_{23} = 0.425 \, , ~~~
\sin^2\theta_{31} = 0.0214 \, , ~~~
\delta_{CP} = 1.38\pi \, .
}
The $CP$-violating physics in our study depends mainly on the Dirac phase and the additional GUT-scale phases.
We will, generally, choose the GUT-scale phases to maximize the $CP$-odd observables and ignore the Majorana phases.
We will give the details of how we determine the GUT-scale phases in the next section.

We also assume a normal mass ordering for the neutrinos, and we use the central values for the mass differences of
\eqs{
m_2^2 - m_1^2 = 7.37 \times 10^{-5} ~\text{eV}^2 \, , ~~~
|\Del m^2| = 2.56 \times 10^{-3} ~\text{eV}^2 \, ,
}
with $\Del m^2 = m_3^2 - (m_2^2-m_1^2)/2$.

\subsection{Soft-Supersymmetry Breaking Parameters}

In our model, the boundary conditions for the soft-supersymmetry breaking parameters are set above the GUT scale.
The following is the soft-supersymmetry breaking Lagrangian in terms of $SU(5)$ multiplets,
\eqs{
- \Lcal_{\mathrm{soft}} & =
(m_{10}^2)_i^j \widetilde \psi^{\ast i} \widetilde \psi_j
+ (m_{\ovl 5}^2)_i^j \widetilde \phi^{\ast i} \widetilde \phi_j
+ (m_N^2)_i^j \widetilde{\ovl n}^{\ast i} \widetilde{\ovl n}_j
+ m_H^2 H^\dag H + m_{\ovl H}^2 \ovl H^\dag \ovl H + m_\Si^2 \Si^\dag \Si\\
&
+ \left(\frac{(A_{10})_{ij}}{4} \widetilde \psi_i \widetilde \psi_j H
+ \sqrt 2 (A_{\ovl 5})_{ij} \widetilde \psi_i \widetilde \phi_j \ovl H
+ (A_N)_{ij} \widetilde \phi_i \widetilde{\ovl n}_j H
+ (B_N)_{ij} \widetilde {\ovl n}_i \widetilde{\ovl n}_j + \text{h.c.} \right) \\
& + \left(
\frac12 M_5 \lam^a \lam^a
+ B_H \mu_H \ovl H H
+ A_\lam \lam \ovl H \Si H
+ \frac{B_\Si \mu_\Si}{2} \Tr \Si^2
+ \frac{A_\Si \lam_\Si}{3} \Tr \Si^3 + \text{h.c.}
\right) \, .
}
Here, $\widetilde \psi\,,\widetilde \phi$, and $\widetilde{\ovl n}$ are the scalar components of $\Psi\,,\Phi$, and $\ovl N$, respectively.
$\lam^a$ are the $SU(5)$ gauginos, and we use the same notation for the scalar components of the Higgs superfields, i.e. $H \,, \ovl H$, and $\Si$.

The boundary conditions for the soft-supersymmetry breaking parameters are set at an initial scale of $M_\ast$.
We impose the following universality conditions on the soft parameters at $M_\ast$,
\eqs{
(m_{10}^2)_i^j& = (m_{\ovl 5}^2)_i^j = (m_N^2)_i^j = m_0^2 \del_i^j \, , \\
m_H^2 & = m_{\ovl H}^2 = m_\Si^2 = m_0^2 \, , \\
(A_{10})_{ij} & = A_0 f^u_{ij} \, , ~~
(A_{\ovl 5})_{ij} = A_0 f^d_{ij} \, , ~~
(A_N)_{ij} = A_0 f^\nu_{ij} \, ,  \\
A_\lam & = A_\Si = A_0 \, , \\
}
and the gaugino mass is set to be $M_5(M_\ast) = M_{1/2}$.

In the next subsection, we give the GUT scale matching conditions.
We use one-loop RGEs for the soft parameters from $M_\ast$ to $\MGUT$, which can be found in Refs.~\cite{Baek:2001kh,Borzumati:2009hu} for the minimal SUSY $SU(5)$ with right-handed neutrinos.
During the RGE evolution, the soft-supersymmetry breaking parameters are affected by the Yukawa couplings of the SU(5) GUT theory.
In particular, the large mixing angles and phases of the neutrino Yukawa matrix lead to large off-diagonal components of the soft mass matrices $m_{\ovl 5}^2$ and $m_N^2$.
These effects result in flavor-changing and $CP$-violating processes, which we discuss in the next section.

\subsection{Matching Conditions at GUT Scale}
Here, we give the GUT scale matching conditions. We begin by discussing the matching conditions for dimensionless couplings.

We use the one-loop matching conditions for the SM gauge couplings in the \DRbar scheme.
\eqs{
\frac{1}{g_1^2(Q)} & = \frac{1}{g_5^2(Q)} - \frac{1}{8 \pi^2} \left[ - 10 \ln \frac{M_X}{Q} + \frac{2}{5} \ln \frac{M_{H_C}}{Q} \right] - \frac{8cv}{\MPl} \, , \\
\frac{1}{g_2^2(Q)} & = \frac{1}{g_5^2(Q)} - \frac{1}{8 \pi^2} \left[ - 6 \ln \frac{M_X}{Q} + 2 \ln \frac{M_\Si}{Q} \right] - \frac{24cv}{\MPl} \, ,\label{eq:GauMat} \\
\frac{1}{g_3^2(Q)} & = \frac{1}{g_5^2(Q)} - \frac{1}{8 \pi^2} \left[ - 4 \ln \frac{M_X}{Q} + 3 \ln \frac{M_\Si}{Q} + \ln \frac{M_{H_C}}{Q} \right] + \frac{16cv}{\MPl} \, . \\
}
Here, $g_1\,,g_2\,,$ and $g_3$ are the gauge couplings of $U(1)_Y \,, SU(2)_L\,,$ and $SU(3)_C$ respectively, and $Q$ is the renormalization scale.
The term proportional to $c$ in \cref{eq:GauMat} is from the higher-dimensional operator containing $\mathcal{W}$ listed in \cref{eq:Pl_supop} and is of order $\Ocal(v/\MPl) \sim 10^{-2}$.
Its contribution to the matching conditions is, therefore, comparable to the one-loop contributions \cite{Tobe:2003yj}.
Since the SM gauge couplings at the GUT scale can be found by renormalization group running, the matching conditions lead to constraints on the GUT scale mass spectrum and couplings \cite{Hisano:1992mh,Ellis:2016tjc},
\eqs{
\frac{3}{g_2^2(Q)} - \frac{2}{g_3^2(Q)} - \frac{1}{g_1^2(Q)} & = \frac{1}{2\pi^2} \frac35 \ln \frac{M_{H_C}}{Q} - 12 \Del \, , \\
\frac{5}{g_1^2(Q)} - \frac{3}{g_2^2(Q)} - \frac{2}{g_3^2(Q)} & = \frac{3}{2\pi^2} \ln \frac{M_X^2 M_\Si}{Q^3} \, , \\
\frac{5}{g_1^2(Q)} + \frac{3}{g_2^2(Q)} - \frac{2}{g_3^2(Q)} & = \frac{6}{g_5^2(Q)} + \frac{15}{2\pi^2} \ln \frac{M_X}{Q} - 18 \Del \, ,
\label{eq:massdetth1}
}
where $\Del = 8 c v/M_{\text{Pl}}$.
The last expression in \cref{eq:massdetth1} can be used to determine the unified gauge coupling $g_5$.

For the case with $\Del = 0$, $M_{H_C}$ and $M_X^2 M_\Si$ are strongly constrained by the low-energy spectrum and couplings, and $M_{H_C}$ should be around $10^{15}~\text{GeV}$ for weak scale supersymmetry breaking.
However, if Planck suppressed operators are taken into consideration, the constraint on $M_{H_C}$ is relaxed considerably \cite{Ellis:2016tjc}.
In fact, the added freedom provided by $\Del$ allows us to treat $\lam$ and $\lam_\Si$ as free parameters, which then control the size of $M_{H_C}$,
\eqs{
M_{H_C}= \lam\left(\frac{M_X^2 M_\Si}{g_5^2 \lam_\Si} \right)^{1/3} \, .
\label{eq:massdetth2}
}
$M_X^2 M_\Si$ is determined by the second expression in \cref{eq:massdetth1}.
The remaining two expressions in \cref{eq:massdetth1} can be written in terms of $\lam\,,\lam_\Si\,,g_5$, and $M_{H_C}$, and so $g_5$ and $M_{H_C}$ are determined by our choice of $\lam $ and $\lam_\Si$.

We use the following tree-level matching conditions for the Yukawa couplings,
\eqs{
y^u_{ij}(Q) & = f^u_{ij}(Q) \, ,  \\
y^d_{ij}(Q) & = f^d_{ij}(Q) + \frac{2v}{\MPl} c_{ij} \, ,  \\
y^e_{ij}(Q) & = f^d_{ji}(Q) - \frac{3v}{\MPl} c_{ji} \, ,  \\
y^\nu_{ij}(Q) & = f^\nu_{ij}(Q) - \frac{3v}{\MPl} c^\prime_{ij} \, ,
\label{eq:yukmatch}}
where $y^u \,, y^d \,, y^e\,,$ and $y^\nu$ are respectively the up-type Yukawa coupling matrix, the down-type Yukawa coupling matrix, the lepton Yukawa coupling matrix, and the neutrino Yukawa coupling matrix in the MSSM with right-handed neutrinos.
$c_{ij}$ and $c^\prime_{ij}$ are the coefficients of the dimension-five operators.
For simplicity, we ignore the dimension-five operators which are irrelevant for $b$-$\tau$ unification.
In our numerical analysis, $c^\prime$ is also negligible since the Majorana masses for the right-handed neutrinos are of order $\Ocal(10^{15})~\mathrm{GeV}$, so that the largest neutrino Yukawa coupling is of order $\Ocal(1)$.

The matching condition for the GUT scale Yukawa coupling matrix $f_d$ is
\eqs{
f_d = \frac35 y_d + \frac25 y_e^T \, ,
}
which is found from the second and third expressions in \cref{eq:yukmatch}.
\footnote{
In regards to the dimension five operators, we can estimate the values from the matching condition \cref{eq:yukmatch}.
The largest coefficient is $c_{33} = 0.04$, and the other components are much less than $c_{33}$.
}
Here, the superscript $T$ denotes the transpose of the matrix.

Next, we consider the matching conditions for the soft-supersymmetry breaking parameters.
When the boundary conditions for the soft-supersymmetry breaking parameters are imposed above the GUT scale, the GUT-breaking VEV has SUSY breaking corrections and is found to be \cite{Hall:1983iz}
\eqs{
\ave{\Si} = \left[ v + \Ocal(\MSUSY) + F_\Si \theta^2 \right] \diag(2,2,2,-3,-3) \, ,
}
to leading order in a typical soft-supersymmetry breaking mass scale $\MSUSY$, where
\eqs{
F_\Si = v (A_\Si - B_\Si) + \cdots \, .
}
Here, ellipses stand for the higher-order correction from soft-supersymmetry breaking parameters of order $\Ocal(M_{\mathrm{SUSY}}^2)$, and thus are only relevant for the $B$-term matching conditions discussed below.
This F-term also contributes to the gaugino masses at the GUT scale via the last term in \cref{eq:Pl_supop}.

Taking into account the one-loop threshold conditions and the Planck-suppressed operator, the matching condition for gaugino masses are given by \cite{Hisano:1993zu,Tobe:2003yj}
\eqs{
\frac{M_1}{g_1^2} & = \frac{M_5}{g_5^2} - \frac{1}{16\pi^2} (10M_5 + 10 A_\Si - 10 B_\Si + \frac{2}{5} B_H) + \frac{\Del(A_\Si - B_\Si)}{2} \, , \\
\frac{M_2}{g_2^2} & = \frac{M_5}{g_5^2} - \frac{1}{16\pi^2} (6M_5 + 6 A_\Si - 4 B_\Si + \frac{2}{5} B_H) + \frac{3\Del(A_\Si - B_\Si)}{2} \, , \\
\frac{M_3}{g_3^2} & = \frac{M_5}{g_5^2} - \frac{1}{16\pi^2} (4M_5 + 4 A_\Si - B_\Si + B_H) - \Del (A_\Si - B_\Si) \, , \\
}
where $\Del$ is defined in \cref{eq:massdetth1}.
$M_1 \,, M_2\,,$ and $M_3$ are the gaugino masses for $U(1)_Y\,,SU(2)_L\,,$ and $SU(3)_C$, respectively.

For the soft scalar mass matrices as well as the scalar-trilinear matrices, we use tree-level matching conditions,
\eqs{
m_{\widetilde Q}^2 = m_{\widetilde u}^2 = m_{\widetilde e}^2 = m_{10}^2 \, ,  & ~~~~~
m_{\widetilde L}^2 = m_{\widetilde d}^2 = m_{\ovl 5}^2 \, , ~~~~~
m_{\widetilde N_R}^2 = m_N^2 \, , \\
m_{H_u}^2 = m_H^2 \, , & ~~~~~
m_{H_d}^2 = m_{\ovl H}^2 \, , \\
(A_u)_{ij} = (A_{10})_{ij} \, , & ~~~~~
(A_d)_{ij} = (A_e)_{ji} = (A_{\ovl 5})_{ij} \, , ~~~~~
(A_\nu)_{ij} = (A_N)_{ij} \, . \\
}
Here, $m_{\widetilde f}^2 ~ (\widetilde f = \widetilde Q\,, \widetilde L\,, \widetilde u\,, \widetilde d\,, \widetilde e, \widetilde N_R)$ denote the $3 \times 3$ soft mass matrices for the sfermions, and $A_f ~ (f = u\,, d\,, e\,,\nu)$ denote the $3 \times 3$ scalar-trilinear matrices.
$m_{H_u}^2$ and $m_{H_d}^2$ are soft mass parameters for the MSSM Higgs doublets, $H_u$ and $H_d$ respectively.

Finally, we comment on the $\mu$ and $B \mu$ terms.
The supersymmetric mass and soft-supersymmetry breaking mass terms for the MSSM Higgs doublets are determined by the EWSB conditions and are found to be
\eqs{
|\mu|^2 & = \frac{m_{H_d}^2 - m_{H_u}^2 \tan^2 \beta + \frac12 m_Z^2 (1-\tan^2\beta) + \Del_\mu^{(1)}}{\tan^2\beta - 1 + \Del_\mu^{(2)}} \,, \\
B\mu & = - \frac12(m_{H_u}^2+m_{H_d}^2+2\mu^2) \sin2\beta + \Del_B \,,
}
where $\Del_B$ and $\Del_\mu^{(1,2)}$ denote loop corrections \cite{deBoer:1994he,Carena:2001fw,Barger:1993gh}.
$\tan\beta = v_u/v_d$ is the ratio of VEVs of the MSSM Higgs doublets.

Matching conditions for the MSSM $\mu$ and $B \mu$ terms in terms of the GUT-scale $\mu_H$ and $B_H \mu_H$ are derived in Ref.~\cite{Borzumati:2009hu}.
These expressions, plus the reality condition on the $A$ and $B$-terms, lead to a non-trivial constraint on the GUT-scale soft parameters given by \cite{Ellis:2016tjc}
\eqs{
A_\Si^2 - \frac{\lam_\Si \mu}{3 \lam} (A_\Si - 4 A_\lam + 4 B) + \left( \frac{\lam_\Si \mu}{6 \lam}  \right)^2 \geq 8 m_\Si^2 \,,
\label{eq:Bcond}
}
which is applied at the GUT scale.
In particular, since we consider the case with $\lam_\Si \ll \lam$, this condition simplifies to $A_\Si^2 \geq 8 m_\Si^2$.
This constraint may be weakened by considering non-zero $CP$ phases for the supersymmetry breaking soft masses. However, in the results presented in \cref{sec:result}, we examine this constraint.

\section{Flavor and CPV Physics \label{sec:FVCPV}}
In the presence of the right-handed neutrinos, the large mixing angles and $CP$ violating phases in the neutrino sector induce large flavor and $CP$ violation in the MSSM soft-supersymmetry breaking parameters.
Proton stability also gives strong constraints on supersymmetric grand unified theories.
In this section, we discuss flavor issues in super-GUTs with right-handed neutrinos, more particularly meson oscillations, lepton flavor violating (LFV) processes, electric dipole moments (EDM), and proton decay.

\subsection{Meson Oscillations}

The flavor and $CP$ violations in the MSSM soft masses arise from RGE effects proportional to the off-diagonal components of $f^\nu$.
This RG running generates off diagonal components of $m_{\widetilde d}^2$ as the soft mass is evolved from $M_\ast$ to the GUT scale.
The interaction which generates this flavor and $CP$ violation involves the colored Higgs boson and is only active above the GUT scale.
An approximate expression for the off-diagonal components of $m_{\widetilde d}^2$ is given by
\eqs{
(m_{\widetilde d}^2)_{ij} & \simeq - \frac{1}{8\pi^2} [f^\nu f^{\nu\dag}]_{ij} (3 m_0^2 + A_0^2) \ln \frac{M_\ast}{\MGUT} \, , ~~~~~
(i \neq j) \, . \\
& \simeq - \frac{1}{8\pi^2} e^{i (\vph_{d_i} - \vph_{d_j})} U_{ik} (f^\nu_k)^2 U^\ast_{jk} (3 m_0^2 + A_0^2) \ln \frac{M_\ast}{M_{\text{GUT}}} \, .
\label{eq:offdiag_dsq}
}
Since neutrino oscillation data indicate that there are large mixing angles and $CP$ phases in the neutrino sector, a large hierarchy between $\MGUT$ and the input scale for the soft mass, $M_\ast$, will lead to large amounts of flavor and $CP$ violation in the right-handed down squark mass matrix.
Flavor and $CP$ violations are also generated in the other squark mass matrices.
However, for the left-handed squark and right-handed up squark mass matrices, the off-diagonal elements are proportional to CKM matrix elements. Because of this, the flavor and $CP$ violations in the other mass matrices, $m_{\widetilde u}^2$ and $m_{\widetilde Q}^2$, are much less significant.
The slepton doublets mass matrix is identical to the down-squark mass matrix at the GUT scale.
However, there are additional flavor and $CP$ violating contributions to the slepton mass matrix from the RG evolution from $M_{GUT}$ down to the scale of the right-handed neutrino masses.
We will discuss the associated lepton flavor issues in the next subsection.
We will refer to the flavor violation from the off-diagonal elements of $m_{\widetilde d}^2$ and $m_{\widetilde L}^2$ as the non-minimal flavor violating (NMFV) contribution.

For quark flavor violation, the strongest constraint comes from meson oscillation measurements, such as $K^0\text{-}\ovl{K}^0$ oscillation.
In general, the meson oscillations arise from the following four-Fermi effective Hamiltonian,
\eqs{
\mathcal{H}^{\Del F = 2}_{\text{eff.}} = \sum_{f = L, R} \sum_{n=1}^3 C_{nf} Q_{nf} + \sum_{n=4}^5 C_n Q_n \, ,
}
where
\eqs{
(Q_{1R})_{ij} & = \ovl d^\alpha_{iR} \gamma_\mu d_{jR}^\alpha \ovl d^\beta_{iR} \gamma^\mu d_{jR}^\beta \, , \\
(Q_{2R})_{ij} & = \ovl d^\alpha_{iL} d_{jR}^\alpha \ovl d^\beta_{iL} d_{jR}^\beta \, , ~~~~~
(Q_{3R})_{ij} = \ovl d^\alpha_{iL} d_{jR}^\beta \ovl d^\beta_{iL} d_{jR}^\alpha \, , \\
(Q_4)_{ij} & = \ovl d^\alpha_{iL} d_{jR}^\alpha \ovl d^\beta_{iR} d_{jL}^\beta \, , ~~~~~
(Q_5)_{ij} = \ovl d^\alpha_{iL} d_{jR}^\beta \ovl d^\beta_{iR} d_{jL}^\alpha \, , \\
}
with color indices $\alpha, \beta$ and flavor indices $i, j$ which are implicit in the effective Hamiltonian.
$C_{nL,R} ~ (n = 1, 2, 3)$ and $C_n ~ (n = 4, 5)$ are the corresponding Wilson coefficients.
The operators $Q_{iL} ~ (i = 1, 2, 3)$ are obtained from $Q_{iR} ~ (i = 1, 2, 3)$ by replacing $R \leftrightarrow L$.

In regards to the SUSY contribution to meson oscillations, we use the mass insertion approximation.
The expression for the relevant Wilson coefficients are \cite{Gabbiani:1996hi}
\eqs{
\left[ C_{1R} \right]_{ij} & = - \frac{\alpha_S^2}{36 m_{\widetilde q}^2} \left( \Del_{ij}^{(R)} \right)^2 \left[ 4 x f_6(x) + 11 \hat f_6(x) \right] \, , \\
\left[ C_4 \right]_{ij} & = - \frac{\alpha_S^2}{3 m_{\widetilde q}^2} \Del_{ij}^{(L)} \Del_{ij}^{(R)} \left[ 7 x f_6(x) - \hat f_6(x) \right] \, , \\
\left[ C_5 \right]_{ij} & = - \frac{\alpha_S^2}{9 m_{\widetilde q}^2} \Del_{ij}^{(L)} \Del_{ij}^{(R)}  \left[ x f_6(x) + 5 \hat f_6(x) \right] \, ,
\label{eq:Wilson_KK}
}
and $C_{1L}$ is obtained from $C_{1R}$ with the exchange $R \leftrightarrow L$.
\footnote{
The other Wilson coefficients, $C_{2L,R}$ and $C_{3L,R}$, are proportional to the left-right squark mixings.
Although the left-right squark mixing terms are omitted from $C_4$ and $C_5$, we have include these contributions in our numerical studies.
}
Here, $m_{\widetilde q}$ is the averaged squark mass, $x = m_{\widetilde g}^2/ m_{\widetilde q}^2$ and $m_{\widetilde g}$ is the gluino mass.
The mass insertion parameters $\Del^{(L,R)}_{ij}$ are defined by
\eqs{
\Del^{(L)}_{ij} \equiv \frac{(m^2_{\widetilde Q})_{ij}}{m_{\widetilde q}^2} \, , ~~~~~
\Del^{(R)}_{ij} \equiv \frac{(m^2_{\widetilde d})_{ij}}{m_{\widetilde q}^2} \, ,
}
and $f_6(x)$ and $\hat f_6(x)$ are the functions arising from the loop calculations.
The Wilson coefficients are computed at the sparticle mass scale.
In our study, we include the two-loop QCD RGE corrections to the Wilson coefficients derived in Ref.~\cite{Buras:2000if}, and  use hadron matrix elements calculated in lattice QCD simulations.
For the $K_0\text{-}\ovl K_0$ mixing, the matrix elements are given by
\eqs{
\evalue{\ovl K^0}{K^0}{Q_{1 L, R}(Q)} & = \frac13 m_K f_K^2 B_1(Q) \, , \\
\evalue{\ovl K^0}{K^0}{Q_{2 L, R}(Q)} & = -\frac{5}{24} \left(\frac{m_K}{m_s+m_d} \right)^2 m_K f_K^2 B_2(Q) \, , \\
\evalue{\ovl K^0}{K^0}{Q_{3 L, R}(Q)} & = \frac{1}{24} \left(\frac{m_K}{m_s+m_d} \right)^2 m_K f_K^2 B_3(Q) \, , \\
\evalue{\ovl K^0}{K^0}{Q_{4}(Q)} & = \frac14 \left(\frac{m_K}{m_s+m_d} \right)^2  m_K f_K^2 B_4(Q) \, , \\
\evalue{\ovl K^0}{K^0}{Q_{5}(Q)} & = \frac{1}{12} \left(\frac{m_K}{m_s+m_d} \right)^2 m_K f_K^2 B_5(Q) \, . \\
}
Here, $m_K$ and $f_K$ are the mass of kaon and the kaon decay constant, respectively.
$m_d$ and $m_s$ are the bare masses for the strange and down quarks, respectively.
$B_i(Q)~(i = 1, \text{-}, 5)$ are referred to as $B$-parameters, and are calculated in lattice simulations.
The numerical values for these $B$-parameters are evaluated at $Q = 2~\text{GeV}$ in the \MSbar scheme, with values:
\eqs{
	B_1(2~\text{GeV}) & = 0.557 \, , ~~~
	B_2(2~\text{GeV}) = 0.568 \, , \\
	B_3(2~\text{GeV}) & = 0.847 \, , ~~~
	B_4(2~\text{GeV}) = 0.984 \, , ~~~
	B_5(2~\text{GeV}) = 0.714 \, ,
}
Here, $B_1(2~\text{GeV})$ is the global fit value of FLAG average \cite{Aoki:2016frl}, and the central values for $B_{\text{2--5}}(2~\text{GeV})$ are found in Ref.~\cite{Jang:2015sla}.
Both of them are evaluated using $N_f = 2 + 1$ flavor QCD lattice simulation.
In the following numerical analysis, we use the lattice results for the hadron matrix elements.

Finally, we define the $K_L$-$K_S$ mass difference $\Del m_K$ and the $CP$-violating parameter $\ep_K$ as follows,
\eqs{
\Del m_K = 2 \text{Re} \evalue{\ovl K^0}{K^0}{\mathcal{H}^{\Del S=2}_{\text{eff.}}} \, , ~~~
\ep_K = \frac{1}{\sqrt 2 \Del m_K}\text{Im} \evalue{\ovl K^0}{K^0}{\mathcal{H}^{\Del S=2}_{\text{eff.}}} \, .
}
The experimental values for these parameters are \cite{Patrignani:2016xqp}
\eqs{
\Del m_K|_\text{exp} & = (3.484 \pm 0.006) \times 10^{-12}~\text{MeV} \, , ~~~ \\
|\ep_K||_\text{exp} & = (2.228 \pm 0.011) \times 10^{-3} \, .
}
The SM value for the $K_L$-$K_S$ mass difference has been calculated using lattice QCD simulation \cite{Bai:2014cva} and the latest SM prediction for $\ep_K$ is found in Ref.~\cite{Ligeti:2016qpi},
\eqs{
\Del m_K|_{\text{SM}} & = 3.19(41)(96) \times 10^{-12}~\text{MeV} \, , ~~~ \\
|\ep_K||_\text{SM} & = 2.24(19) \times 10^{-3} \, .
\label{eq:KKbar_SM}
}
According to Ref.~\cite{Ligeti:2016qpi}, the SUSY contribution to $|\ep_K|$, denoted by $|\ep_K^{\mathrm{SUSY}}|$, should be less than $0.31 |\ep_K|_\text{SM} \simeq 0.69 \times 10^{-3}$.
For the new physics contribution to $\Del m_K$, we will impose that a deviation from $\Del m_K|_{\text{SM}}$ should be within the experimental and theoretical errors.
However, a previous study showed that the SUSY contribution to $\Del m_K|_{\text{SUSY}}$ in the presence of the right-handed neutrinos was much smaller than the SM value \cite{Moroi:2000mr}.

\subsection{Lepton Flavor Violation Processes}
As mentioned in the previous subsection, significant flavor violation in the left-handed slepton sector is induced by the large neutrino mixings.
The off-diagonal elements of the slepton soft mass matrix are approximately given by
\eqs{
(m_{\widetilde L}^2)_{ij} \simeq - \frac{1}{8\pi^2} \sum_k \hat f^\nu_{ik} (\hat f^{\nu\dag})_{kj} (3 m_0^2 + A_0^2) \ln \frac{M_\ast}{(M_R^D)_k}  \, , ~~~~~
(i \neq j) \, .
\label{eq:offdiag_sl}
}
Here, we define $\hat f^\nu_{ij} \equiv f^\nu_{ik} e^{-i\vph_{\nu_k}} W^\ast_{jk} e^{- i\ovl\vph_{\nu_j}}$ in terms of the unitary matrix $W$ which rotates the right-handed neutrino superfields.

The lepton flavor violation (LFV) we consider is  $\mu \to e + \gamma$.
According to a recent study on the LFV processes \cite{Crivellin:2018mqz}, this mode places the strongest constraint on LFV processes.
The latest upper limit on the branching ratio for this process is
\eqs{
Br(\mu \to e \gamma) < 4.2 \times 10^{-13} \, ,
}
which comes from the MEG experiment \cite{TheMEG:2016wtm}.

The lepton LFV processes in supersymmetric models have been discussed in Refs.~\cite{Borzumati:1986qx,Leontaris:1985pq,Hisano:1995nq,Hisano:1995cp}.
A decay rate for $l_i \to l_j \gamma$ process is given by
\eqs{
\Gamma(l_i \to l_j \gamma) = \frac{m_{l_i}^3}{4\pi} |(C_{LL} + C_{LR})_{ji}|^2 \, ,
\label{eq:LFV}
}
where $m_{l_i}$ is the mass of the lepton $l_i$ and $C_{LL}$ and $C_{LR}$ are Wilson coefficients which were derived in Ref.~\cite{Hisano:1995nq,Hisano:1995cp}.
The source of $C_{LL}$ Wilson coefficient is the LFV elements of the soft mass matrix $m_{\widetilde L}^2$.
On the other hand, the $C_{LR}$ Wilson coefficient is from the left-right slepton mixing matrix.
To compute the LFV branching fraction, we use the average value for muon lifetime to determine the total decay rate of the muon; $\Gamma_\mu^{-1} = \tau_\mu = 2.20 \times 10^{-6} \, \mathrm{s}$ \cite{Patrignani:2016xqp}.

\subsection{Electric Dipole Moments \label{sec:edm}}

Other possible signals of flavor and $CP$ violation are electric dipole moments (EDMs).
As already discussed above, the flavor and $CP$ violation of the neutrino sector in the PMNS matrix and GUT scale phase induces flavor and $CP$ violation in the squark and slepton mass matrices.
Although the  soft masses could have other sources of $CP$ violation, we assume real soft masses at the input scale and examine the effects of the induced flavor and $CP$ violation on EDMs.

The electron EDM can be a potential signal of this induced $CP$-violation in the soft-supersymmetry breaking parameters.
The most recent upper bound on the electron EDM, $|d_e| \lesssim 9.3 \times 10^{-29} [e~\text{cm}]$ \cite{Baron:2016obh}, provides the most stringent constraint on our model, as will be seen below.

Although the electron EDM places the most stringent constraint on our model, we have also calculated the quark EDMs and quark chromo-EDMs (CEDMs).
The EDMs and CEDMs are calculated at the SUSY scale.
We include both the one-loop and two-loop contributions (see Refs.~\cite{Giudice:2005rz,Ellis:2008zy,Li:2008kz}) to the parton-level EDMs in our numerical studies, however, the bino (neutralino)-sfermion one-loop diagrams with sfermion flavor violation give the dominant contribution to the EDMs.
Due to a chirality flip in the loop diagram for the dipole, the diagrams with loops containing the heavier fermions provide the dominant contribution, if flavor mixing is sufficiently large.
The neutralino one-loop contribution with flavor violation to the electron EDM is approximately given by \cite{Hisano:2008hn}
\eqs{
\frac{d_e}{e} \sim \frac{g_Y^2}{32\pi^2} \frac{m_\tau}{m_{\widetilde l}^2} \frac{\mu M_1}{m_{\widetilde l}^2}\text{Im}[(\Del^{(L)}_l)_{13} (\Del^{(R)}_{l})_{31}] f(x) \,,
}
where $f(x)$ is a loop function and $x = M_1^2/m_{\widetilde l}^2$, with $m_{\widetilde l}$ the averaged slepton mass.
$M_1$ is the mass of binos and $\mu$ is the supersymmetric mass parameter for Higgsinos.
The mass insertion parameters are given by
\eqs{
(\Del^{(L)}_l)_{ij} \equiv \frac{(m_{\widetilde L}^2)_{ij}}{m_{\widetilde l}^2} \, , ~~~~~
(\Del^{(R)}_l)_{ij} \equiv \frac{(m_{\widetilde e}^2)_{ij}}{m_{\widetilde l}^2} \, .
}

For the hadronic EDMs, we first evaluate the quark-level EDMs at the SUSY scale.
We then RG evolve the Wilson coefficients, including the mixing effects of $CP$-odd operators  \cite{Degrassi:2005zd}, down to $Q=1~\text{GeV}$.
The nucleon EDMs at the hadronic scale $Q = 1~\text{GeV}$ are found from the following relation, determined by lattice simulations and QCD sum rules \cite{Hisano:2012sc,Hisano:2015rna},
\eqs{
d_p & = 0.78 d_u - 0.20 d_d + e(-1.2 \widetilde d_u - 0.15 \widetilde d_d) \, , \\
d_n & = - 0.20 d_u + 0.78 d_d + e(0.29 \widetilde d_u + 0.59 \widetilde d_d) \, , \\
}
Here, $d_p$ and $d_n$ are the proton and neutron EDMs respectively, $d_q$ and $\widetilde d_q$ $(q = u, d)$ are the quark EDMs and CEDMs respectively.

\subsection{Proton Decay}

In supersymmetric $SU(5)$ GUTs, the dominant decay mode for the proton is controlled by the dimension-five operators generated by color-triplet Higgs boson exchange.
The $p \to K^+ \ovl\nu$ mode is the main decay mode induced by these operators, and the latest constraint on this partial decay mode is $\tau(p \to K^+\ovl \nu) \gtrsim 6.6 \times 10^{33}~\mathrm{years}$ \cite{Abe:2014mwa,Takhistov:2016eqm}.
Although the other decay modes, $p\to \pi^+\ovl \nu$ and $n \to \pi^0 \ovl \nu$, are generated by the same operators, these decay modes are suppressed by CKM mixings angles.
Furthermore, the experimental bounds on these decay modes are much weaker than the $p \to K^+ \ovl\nu$ mode \cite{Ellis:2016tjc}.
Therefore, we consider only the $p \to K^+ \ovl\nu$ mode in this paper. Here, we discuss the relevant aspects of proton decay for our model.
For details of this calculation, see Refs.~\cite{Hisano:1992jj,Hisano:2013exa,Nagata:2013sba,Ellis:2015rya,Evans:2015bxa}.

The effective Lagrangian, which controls proton decay, is found by integrating out the color-triplet Higgs boson and is given by
\eqs{
\Lcal_{\text{eff.}}^{\Del B=1} & =  C_{5L}^{ijkl} \Ocal^{5L}_{ijkl} + C_{5R}^{ijkl} \Ocal^{5R}_{ijkl} + \mathrm{h.c.} \,,
}
with
\eqs{
\Ocal^{5L}_{ijkl} & \equiv \int d^2 \theta ~ \frac12 \ep_{\alpha\beta\gamma} (Q^\alpha_i Q^\beta_j) (Q^\gamma_k L_l) \, , \\
\Ocal^{5R}_{ijkl} & \equiv \int d^2 \theta ~ \ep^{\alpha\beta\gamma} \ovl U_{i\alpha} \ovl E_j \ovl U_{k\beta} \ovl D_{l\gamma} \, ,
\label{eq:D5pdecay}
}
where parentheses denote contractions of $SU(2)_L$ indices.
The Greek letters represent $SU(3)_C$ indices, $i,j,k,l$ represent the generations, and $\ep^{\alpha\beta\gamma}$ is the totally antisymmetric tensor.
In the mass basis for the MSSM chiral multiplets, the Wilson coefficients at the GUT scale are given by
\eqs{
C_{5L}^{ijkl} & = - \frac{1}{M_{H_C}} f^u_i e^{i\varphi_{u_i}} \del^{ij} V_{kl}^\ast f^d_l \, , \\
C_{5R}^{ijkl} & = - \frac{1}{M_{H_C}} f^u_i V_{ij} V_{kl}^\ast f^d_l e^{-i\varphi_{u_k}} \, . \\
}
Here, $V_{ij}$ is the CKM matrix, and $\varphi_{u_i}$ are the additional phases in the up-type GUT Yukawa matrix $f^u$.

We determine the proton lifetime by first one-loop RG running the Wilson coefficients $C_{5L}$ and $C_{5R}$ down to the SUSY scale.
The RGEs for these Wilson coefficients, found in Ref.~\cite{Munoz:1986kq}, are modified in the presences of right-handed neutrinos to become
\eqs{
\beta_{5L}^{ijkl} \equiv \mu \diff{1}{}{\mu} C_{5L}^{ijkl} =
\beta_{5L;\text{MSSM}}^{ijkl} + \frac{1}{16\pi^2}(y^\nu y^{\nu\dag})_{lm} C_{5L}^{ijkm} \, ,
}
where $\beta_{5L;\text{MSSM}}^{ijkl}$ denotes the MSSM contribution to the RGE for $C_{5L}^{ijkl}$ \cite{Munoz:1986kq}, and $y^\nu$ is the neutrino Yukawa matrix.
Above the right-handed neutrino mass scale $\beta_{5L}^{ijkl}$ is used in the RGEs and $\beta_{5L;\text{MSSM}}^{ijkl}$ is used below the right-handed neutrino mass scale.
At the SUSY scale, the sfermions in the effective operators in \cref{eq:D5pdecay} are integrated out via charged wino and higgsino exchange processes, giving the four-fermion interactions leading to proton decay.
The large flavor violation in the right-handed down-squark sector does not significantly induce the other modes, such as $p \to \pi^0 \mu^+$ \cite{Nagata:2013sba}.

Since the SUSY scale is a bit higher than the electroweak (EW) scale, we use the SM RGEs to evolve the coefficients from the SUSY scale to the EW scale \cite{Abbott:1980zj}.
Below the EW scale, we evolve the coefficients using the two-loop long-distance corrections \cite{Nihei:1994tx} to obtain the coefficients at the hadronic scale.
The hadron matrix elements are then evaluated at the hadronic scale using lattice simulation \cite{Aoki:2017puj},
\eqs{
\evalue{K^+}{p}{(us)_R d_L} & = ~ - 0.049 \text{GeV}^2 \, , ~~~~~
\evalue{K^+}{p}{(us)_L d_L} = ~ 0.041 \text{GeV}^2 \, , \\
\evalue{K^+}{p}{(ud)_R s_L} & = ~ - 0.134 \text{GeV}^2 \, , ~~~~~
\evalue{K^+}{p}{(ud)_L s_R} = ~ 0.139 \text{GeV}^2 \, . \\
}

\section{Results \label{sec:result}}

Before showing our results, we define our parameters.
The GUT scale, $\MGUT$, is defined as the scale where the condition $g_1(\MGUT) = g_2(\MGUT)$ is satisfied.
The unified coupling $g_5$ at $\MGUT$ is then determined using the third relation in \cref{eq:massdetth1}.
The SUSY scale is defined as the geometric mean of the stop mass eigenvalues, $\MSUSY = \sqrt{m_{\widetilde t_1} m_{\widetilde t_2}}$.
To ensure longevity of the proton, we also take the GUT scale Yukawa couplings to be $\lam(\MGUT) = 0.5$ and $\lam_\Si(\MGUT) = 10^{-4}$ \cite{Ellis:2016tjc}.

The input parameters of our model are then
\eqs{
m_0^2 \, , ~ M_{1/2} \, , ~ A_0 \, , ~ \tan\beta \, , ~ \mathrm{sgn}(\mu) \, , ~ M_\ast \, , ~ (M_R)_{ij} \, ,
~ \vph_{u_i} \, , ~ \vph_{d_i} \, , ~ \vph_{\nu_i} \, , ~ \ovl \vph_{\nu_i} \, ,
}
after fixing $\lam$ and $\lam_\Si$.
For simplicity, we assume the right-handed neutrino mass matrix is proportional to unity, $(M_R)_{ij} = M_{N_R} \del_{ij}$, and is real.
Consequently, we find $\ovl \vph_{\nu_1} = \ovl \vph_{\nu_2} = \ovl \vph_{\nu_3} = 0$ and $\vph_{\nu_1} = \vph_{\nu_2} = \vph_{\nu_3} = 0$.
The input scale for the soft-supersymmetry breaking parameters, $M_\ast$, is set to the reduced Planck mass $\MPl = 2.4 \times 10^{18}~\text{GeV}$.
This is because we want to determine what are the most stringent constraints possible from the flavor and $CP$ violation of the neutrino sector.

The GUT-scale phases $\vph_u$ are chosen to maximize the lifetime of the proton.
This allows us to focus on the constraints coming from flavor and $CP$ violation.
However, even with this maximization procedure, $\tan \beta\lesssim 6$ is required to get a sufficiently long proton lifetime. We take $\tan\beta=6$ for our study.
The remaining phases, $\vph_d$, are determine by maximizing $|\ep_K^\mathrm{SUSY}|$ and the electron EDM.

\begin{figure}
	\centering
	\includegraphics[width=7cm,clip]{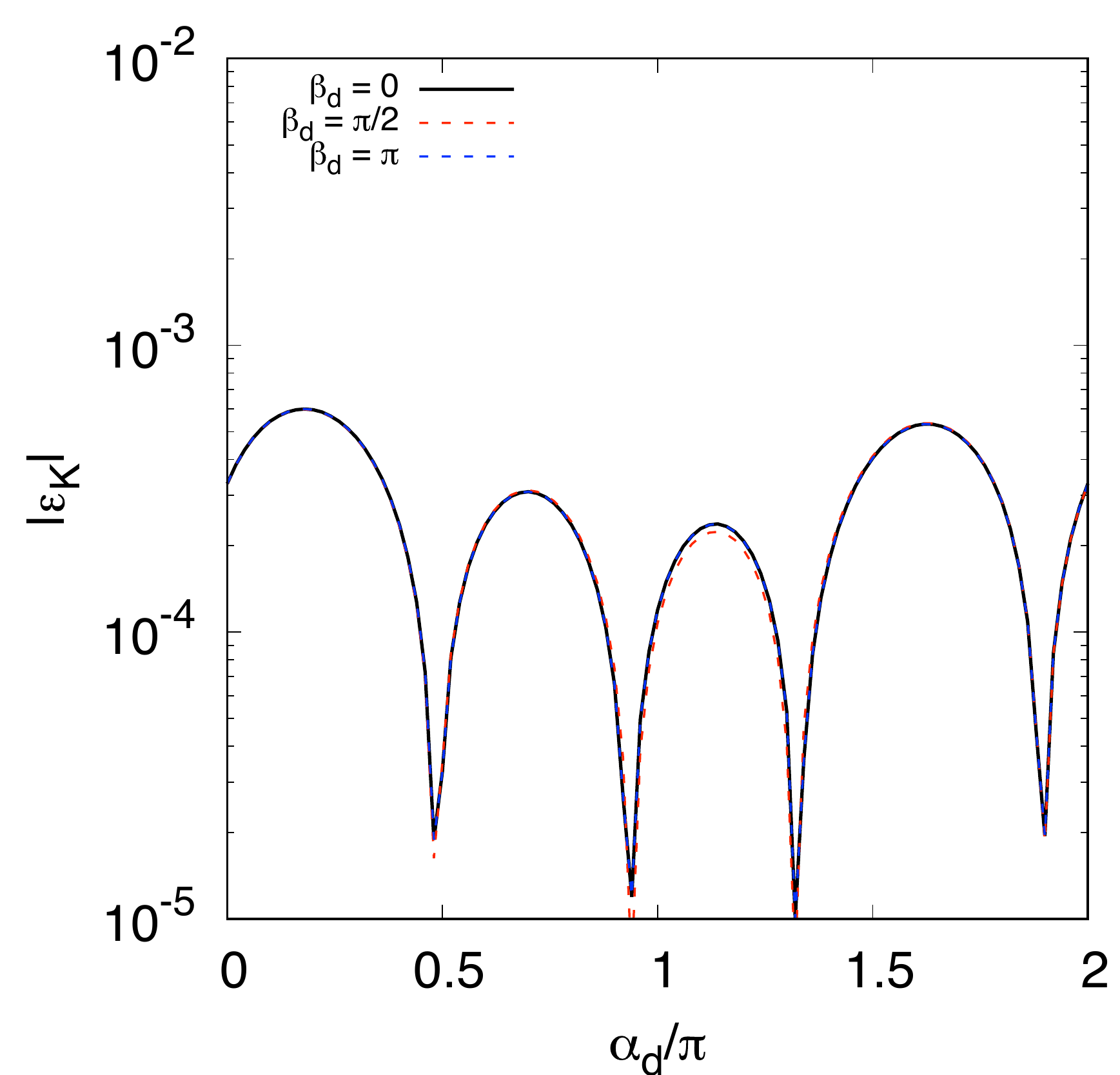} ~~
	\includegraphics[width=7cm,clip]{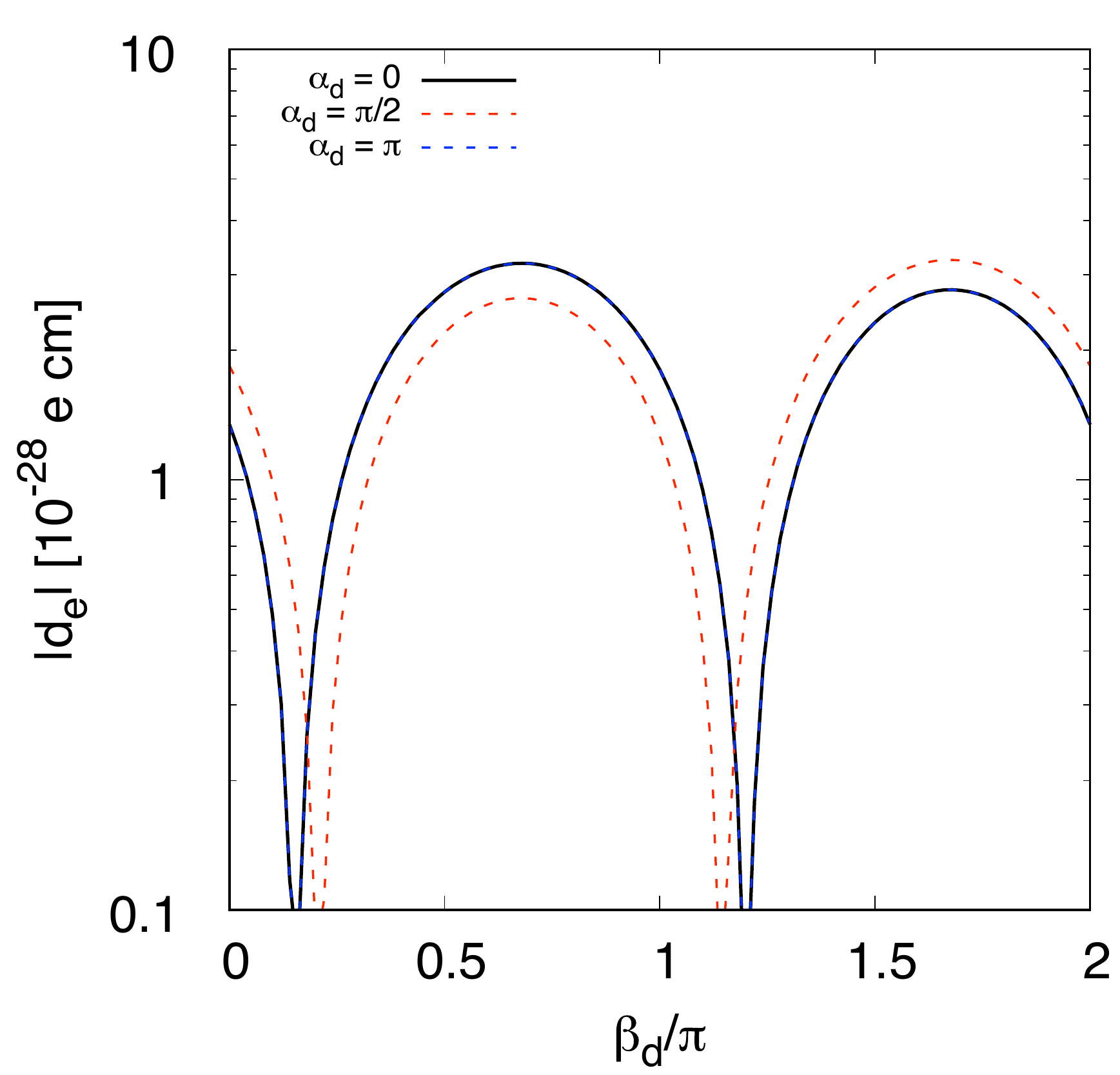}
	\caption{GUT-scale phase dependence of $|\ep_K|$ and $d_e$.
	Different lines correspond to different parameter choices.
	$\tan\beta = 6$, $M_{1/2} = 1~\text{TeV}$, $A_0 = 0$, $\mu > 0$, and $M_\ast = \MPl$ are assumed.
	$|\ep_K^\mathrm{SUSY}|$ is maximized at $\alpha_d = 0.18 \pi$, and
	$|d_e|$ is maximized at $\beta_d = 0.68 \pi$.
	}
	\label{fig:phasedep}
\end{figure}

Because the phases are constrained, $\vph_{d_1} + \vph_{d_2} + \vph_{d_3} = 0$, there are only two independent phases which we will denote by $\alpha_d \equiv \vph_{d_1} - \vph_{d_2}$ and $\beta_d \equiv \vph_{d_1} - \vph_{d_3}$.
$\ep_K$ is most strongly dependent on $(m^2_{\widetilde d})_{12}$ and $\alpha_d$. $\alpha_d$ is then determined by maximizing $\ep_K$.

In the left panel of \cref{fig:phasedep}, we show the $\alpha_d$-dependence of $|\ep_K^\mathrm{SUSY}|$ with $M_{N_R} = 10^{15}~\text{GeV}$.
We set the initial values of the soft-supersymmetry breaking parameters to be $M_{1/2} = 1~\text{TeV} \,, m_0 = 1~\text{TeV} \,, A_0 = 0 \,, \tan \beta = 6 \,,$ and $ \mathrm{sign}(\mu) > 0$.
Because $C_{1R}$ defined in \cref{eq:Wilson_KK} is proportional to $(m^2_{\widetilde d})_{12}^2$, four peaks appear in the plot for $|\ep_K|$.
Since the flavor and $CP$ violation in the left-handed squark sector is so much smaller than that in the right-handed sector, the $|\ep_K^\mathrm{SUSY}|$ is almost completely controlled by the NMFV contribution to the right-handed squark sector.
The positions of the peaks are determined by the $CP$ violation in the neutrino sector and are, therefore, mostly independent of the initial values of the soft-supersymmetry breaking parameters.
In the left panel of \cref{fig:phasedep}, we also show $|\ep_K^{\mathrm{SUSY}}|$ for different values of $\beta_d$.
As is seen in the figure, $|\ep_K^{\mathrm{SUSY}}|$ is almost completely independent of $\beta_d$.
The maximum value for  $|\ep_K^{\mathrm{SUSY}}|$ is found for $\alpha_d \simeq 0.18 \pi$.

As mentioned in the previous section, the dominant contribution to the electron EDM is from the neutralino-stau loop diagram.
This dominant contribution is proportional to $(m^2_{\widetilde L})_{13}$, and is therefore controlled by $\beta_d$.
In the right panel of \cref{fig:phasedep}, we show the $\beta_d$-dependence of $d_e$, with the same input values for the soft-supersymmetry breaking parameters as the left panel.
As is clear from the right panel of \cref{fig:phasedep}, $d_e$ is only weakly dependent on $\alpha_d$ and is maximized for $\beta_d \simeq 0.68 \pi$.
Although $d_e$ is only weakly dependent on $\alpha_d$, this phase can be important in regions with huge cancellations.
For instance, the two-loop stop-associated Barr-Zee contribution has a different $CP$ phase dependence and can, therefore, have the opposite sign of the one-loop contribution. In some of the figures below, we will see that the cancellation between these two contributions can be important.

\begin{figure}
	\centering
	\includegraphics[width=7cm,clip]{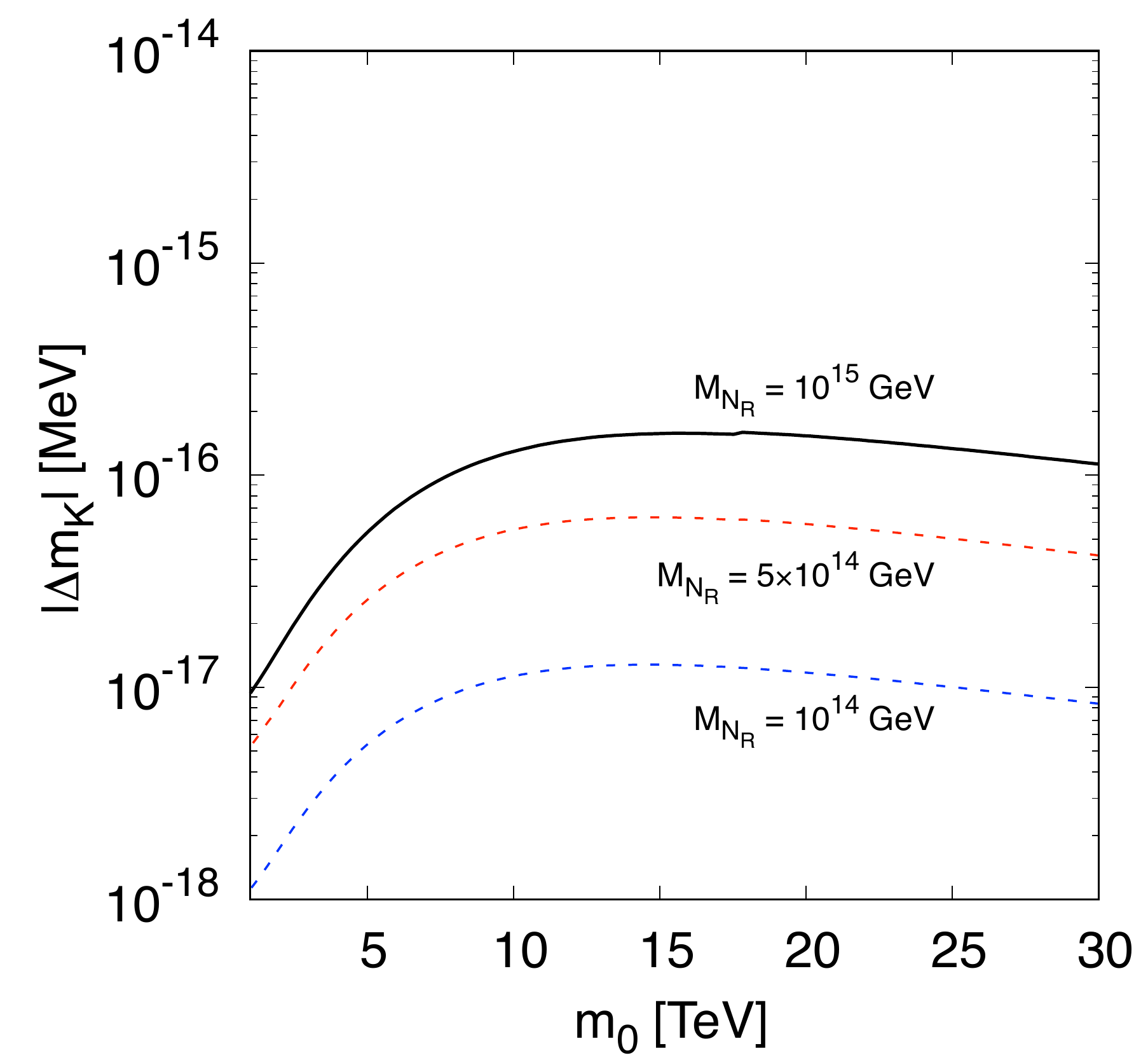}
	\includegraphics[width=7cm,clip]{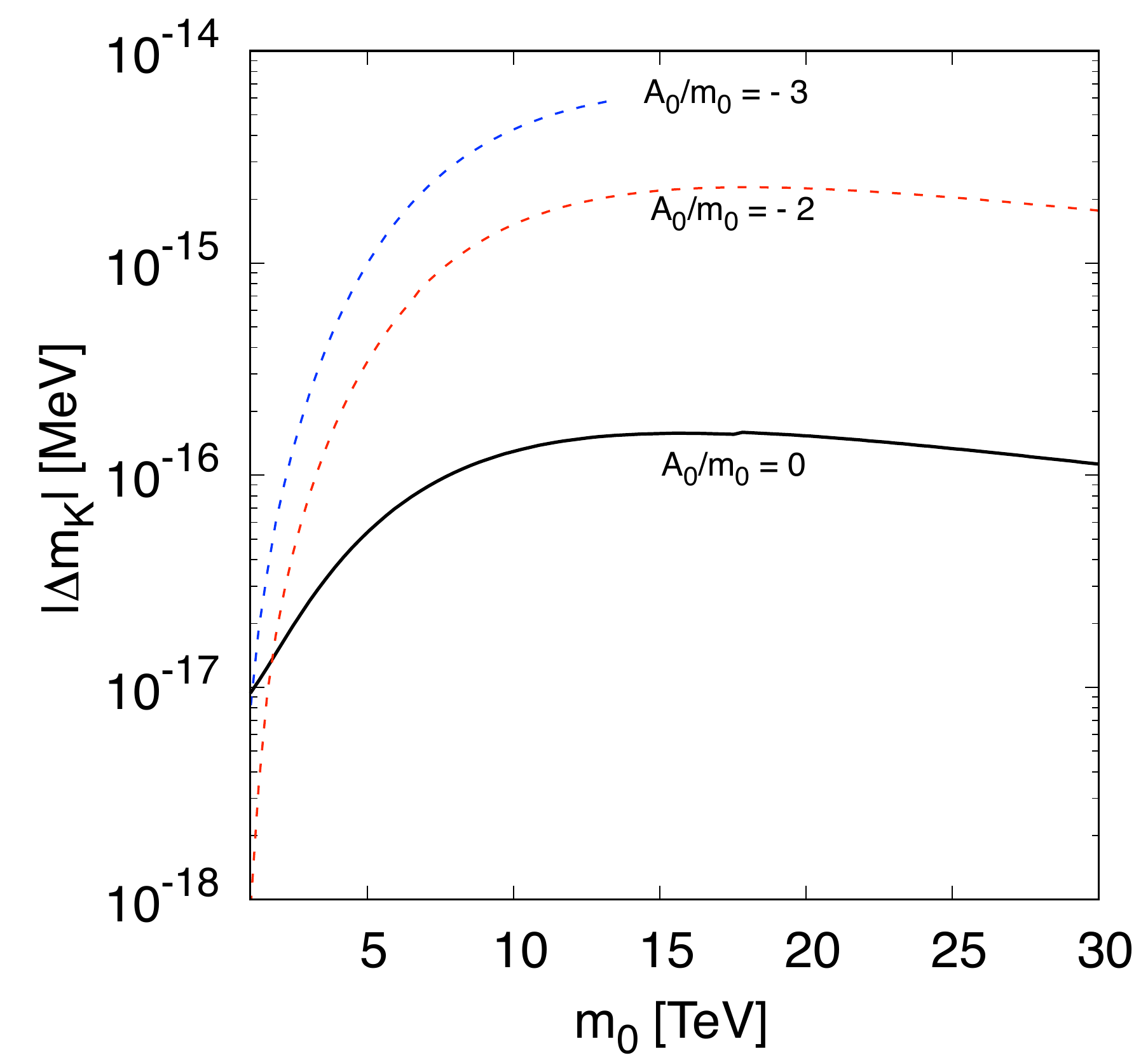}
	\caption{SUSY contribution to kaon mass difference $|\Del m_K|_{\text{SUSY}}$ as a function of $m_0$.
	$\tan\beta = 6$, $M_{1/2} = 5~\text{TeV}$, $\mu > 0$, and $M_\ast = M_{\text{Pl}}$ are assumed.
	$A_0 = 0$ in the left panel, while universal right-handed neutrino mass $M_{N_R}$ is fixed to be $10^{15}~\text{GeV}$ in the right panel.
	}
	\label{fig:DKvsm0}
\end{figure}

Before we show the $m_0\text{-}M_{1/2}$ and $m_0\text{-}A_0$ planes of this model, we consider the dependence of $\Del m_K$, the kaon mass difference, on the SUSY breaking parameters.
The mass difference $\Del m_K$ can be decomposed into two pieces, $\Del m_K = \Del m_K|_{\text{SM}} + \Del m_K|_{\text{SUSY}}$, with $\Del m_K|_{\text{SM}}$ corresponding to the SM value \cref{eq:KKbar_SM}.
The SUSY contribution, $\Del m_K|_{\text{SUSY}}$, should not be larger than the experimental and the theoretical uncertainties.
\cref{fig:DKvsm0} shows the $m_0$-dependence of $\Del m_K|_{\text{SUSY}}$ for various parameter choices of $A_0$ and $M_{N_R}$ as a function of $m_0$.
For small $m_0$, the SUSY contribution to $\Del m_K$ is suppressed because the dominant contribution to the squark mass is from the gaugino radiative correction.

For the left panel in \cref{fig:DKvsm0}, we vary the universal Majorana neutrino mass as $M_{N_R} = 10^{15}~\text{GeV} \,, 5 \times 10^{14}~\text{GeV}\,,$ and $10^{14}~\text{GeV}$ from top to bottom, with fixed $A_0 = 0$.
The $M_{N_R}$ dependence of $|\Del m_K|$ is explained by the neutrino Yukawa dependence on $M_{N_R}$.
The SUSY contribution to $|\Del m_K|$ is largest for large $M_{N_R}$, because the neutrino Yukawa couplings are larger leading to more flavor violation.

In the right panel of \cref{fig:DKvsm0}, we show the $A_0$-dependence of $\Del m_K|_{\text{SUSY}}$ for $M_{N_R} = 10^{15}~\text{GeV}$.
The flavor violation in the down-squark sector induced by RGE running from $M_\ast$ to $\MGUT$ has a piece proportional to an A-term.
This dependence leads to the growth of $|\Del m_K|$ as $|A_0/m_0|$ is increased.
However, if the $A$-terms are taken too large, some of the sfermions become tachyonic.
This is seen in \cref{fig:DKvsm0} by the fact that the line with $A_0/m_0 = - 3$ terminates for $m_0\sim 13$ TeV.

Examining both panels of \cref{fig:DKvsm0}, we see that the SUSY contribution to the kaon mass difference is less than $\Ocal(1)~\%$.
Although $\Del m_K|_{\text{SUSY}}$ is much smaller than the SM predictions, we include it since it is important for calculating $\ep_K^{\mathrm{SUSY}}$ in what follows.

\begin{figure}
	\centering
	\includegraphics[width=7.5cm,clip]{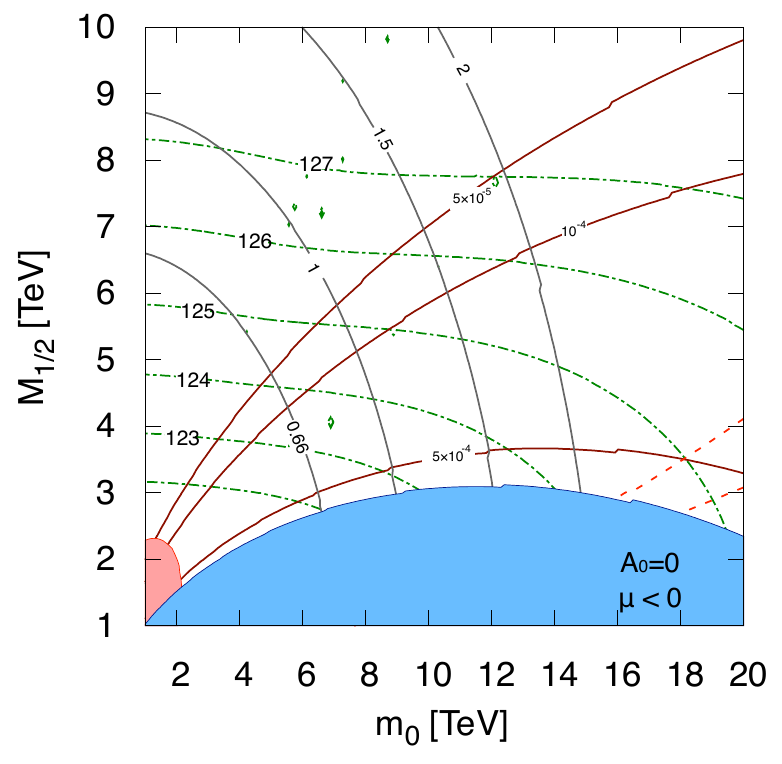}
	\includegraphics[width=7.5cm,clip]{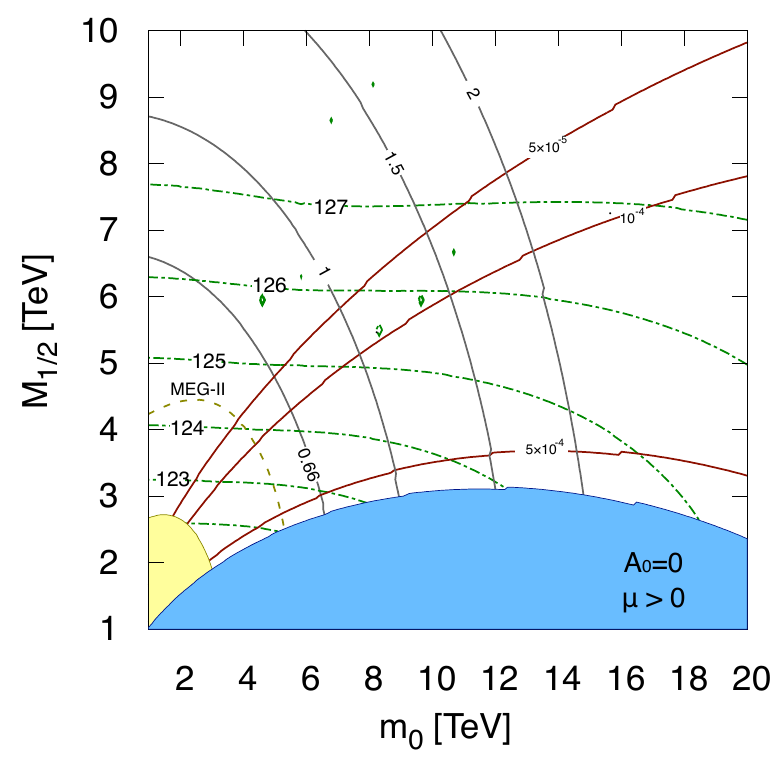} \\
	\includegraphics[width=7.5cm,clip]{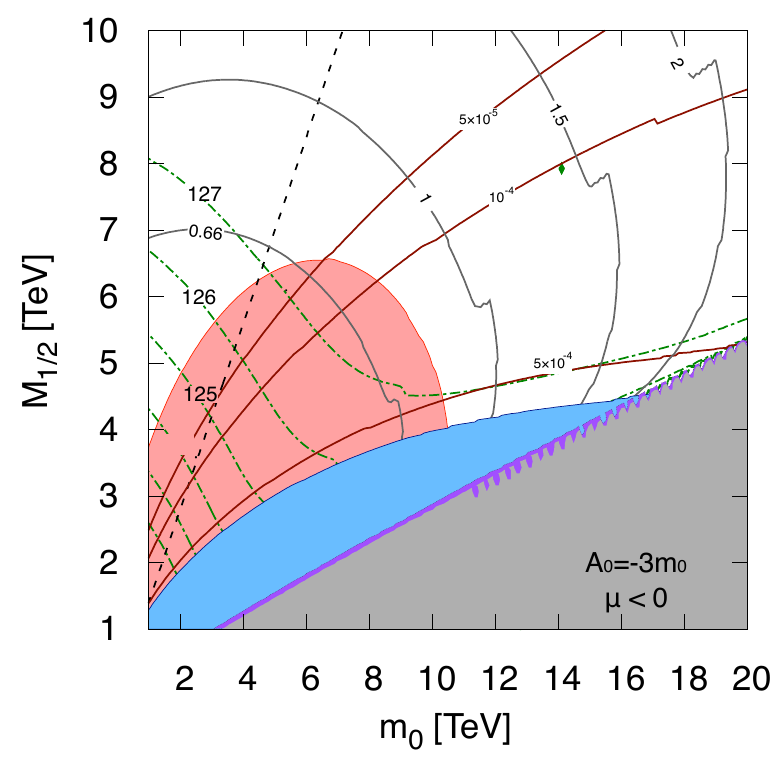}
	\includegraphics[width=7.5cm,clip]{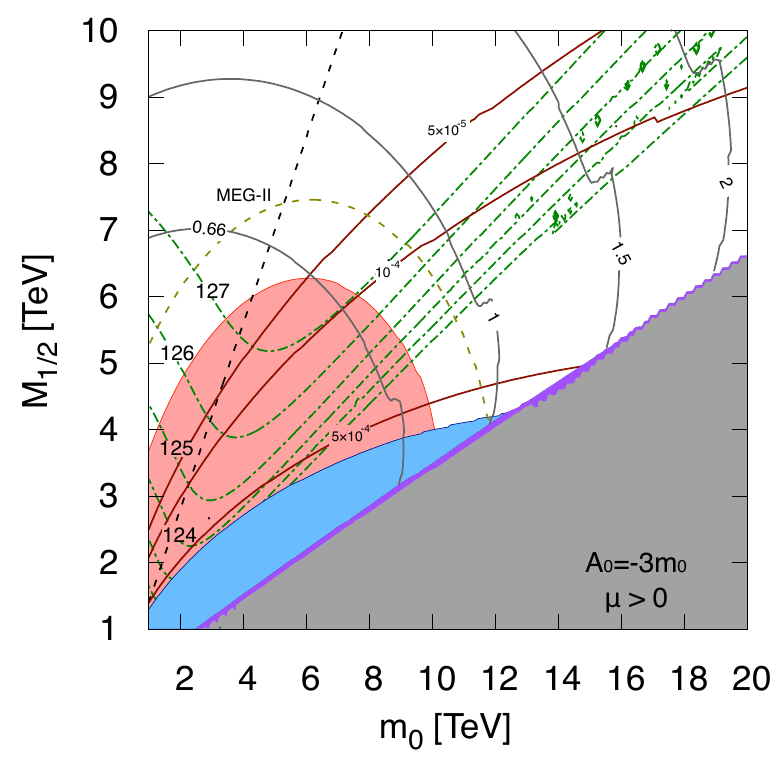}
	\caption{The $m_0$-$M_{1/2}$ plane for
	$\tan\beta = 6$\,, $M_\ast = \MPl$\,, $M_{N_R} = 10^{15}~\mathrm{GeV}\,$, and $\mu < 0$ ($\mu > 0$) in the left (right) panels.
	$A_0 = 0 \, (-3m_0)$ is assumed in the top (bottom) panels.
	The red-brown lines show the SUSY contribution to $|\ep_K|$, with the contribution exceeding $0.69 \times 10^{-3}$ in the cyan-shaded region.
	The black-solid lines indicate the partial proton lifetime $\tau(p \to K^+ \ovl \nu)$ in units of $10^{34}$ years.
	The green dotted lines illustrate the mass of the light Higgs boson.
	The red and yellow-shaded region are excluded by the current electron EDM and $\mu \to e\gamma$ bound respectively.
	The mass difference $m_{\widetilde t_1} - m_{\chi^0_1}$ is below $100~\text{GeV}$ in the purple shaded region.
	}
	\label{fig:m0mhalf0}
\end{figure}

The level of precision  and method used to calculate the SUSY spectrum is as follows.
For the soft-supersymmetry breaking parameters we use the one-loop RGEs between $M_\ast$ and the GUT scale and the matching conditions discussed in \cref{sec:model} are used at the GUT scale.
The RGEs for the soft masses of scalars and gauginos below the GUT scale are at the two-loop level and the other parameters are at the one-loop level.
We, of course, take into account the complex phases and the off-diagonal flavor mixing parts of the RGEs, since our focus is on flavor and $CP$ violation.
The mass spectrum and mixing matrices for the Higgs boson, sfermions, neutralinos, and charginos are evaluated using \FH.

Now, in \cref{fig:m0mhalf0} we show the $m_0$-$M_{1/2}$ plane with $ M_\ast = \MPl \,, M_{N_R} = 10^{15}~\mathrm{GeV}\,,$ and $\tan\beta = 6$.
The difference between the left and right panels is just the sign of the $\mu$-parameter; positive (negative) $\mu$ is used in the right (left) panels.
We set different values for $A_0/m_0$ in the top and bottom panels; $A_0 = 0 \, (-3 m_0)$ is assumed in the top (bottom) panels.
In each figures, we have plotted the Higgs mass contours (green dotted-lines), proton decay constraints (black lines), $|\ep_K^{\text{SUSY}}|$ contours (red-brown lines), future prospect for the electron EDM ($\sim 10^{-31}~e~\text{cm}$ shown by the red-dashed lines) and the MEG-II sensitivity to $\mu \to e \gamma$ decays ($Br(\mu \to e \gamma) \sim 6 \times 10^{-14}$ \cite{Baldini:2018nnn} shown by the yellow-dashed line).
In the purple shaded region, the mass difference between the lightest neutralino and stop, $m_{\widetilde t_1} - m_{\chi^0_1}$, is less than $100~\mathrm{GeV}$.
\footnote{
Since the bino mass for the entire purple regions is less than about 3 TeV and the A-terms are large, some of the purple shaded region should have a viable dark matter candidate due to coannihilation of the Bino with the stop \cite{Ellis:2014ipa,Ellis:2018jyl}.
}
The other shaded regions in \cref{fig:m0mhalf0} are excluded because the electron EDM is larger than $9.3\times 10^{-29}~[e~\text{cm}]$ (the red region), the branching fraction for $\mu\to e\gamma$ is larger than $4.2 \times 10^{-13}$ (the yellow region), or $|\ep_K^{\mathrm{SUSY}}|$ exceeds $0.69 \times 10^{-3}$ (the cyan region).
Even though we have checked the nucleon EDMs in this paper, the constraints are much weaker than the constraints presented.
According to \FH, the SUSY contribution to the $Br(B \to X_s \gamma)$ is less than $1\%$ of the SM prediction, and therefore can be ignored.

Since off-diagonal components of soft-masses are proportional to $m_0^2$ and $A_0^2$ (\cref{eq:offdiag_dsq,eq:offdiag_sl}), no flavor and $CP$ violation is generated if $m_0$ vanishes at the input scale.
The flavor and $CP$ violation is maximized when sfermion masses are dominated by $m_0^2$ not $M_{1/2}$, and then the violation decreases when sfermions are decoupled.
Therefore, the flavor and CP violating observables have a peak along $m_0$ axis for fixed $M_{1/2}$.

In the top panels of \cref{fig:m0mhalf0}, the LSP is a bino-like neutralino throughout the entire plane.
The RGE running of the soft-supersymmetry breaking parameters above the GUT scale lifts the mass of all charged scalars above the neutralino mass.
Because of the right-handed neutrinos, the mass spectrum at the GUT scale is altered.
Without the right-handed neutrino, the lightest sfermion mass is $(m_{10}^2)^3_3$, due to RG running effects of the top Yukawa coupling.
However, the neutrino Yukawa couplings suppress the mass of $\Phi_3$ leading to $(m_{10}^2)^3_3> (m_{\ovl 5}^2)^3_3$.

In our setup, due to the large Yukawa couplings for the neutrinos and the large group-theoretic numerical factors of SU(5), the diagonal components of $m_N^2$, the soft-mass matrix for the right-handed neutrinos, is driven negative by RGE effects.
A positive $m_N^2$ drives $m_{\ovl 5}^2$ to smaller values, as seen in Refs.~\cite{Moroi:1993ui,Kadota:2009vq,Kadota:2009fg}.
However, because $m_N^2$ promptly turns negative, the mass of $m_{\ovl 5}^2$ is not so suppressed.

The sign of $\mu$ affects flavor and $CP$ violating processes involving left-right sfermion mixing.
This dependence is rather important when the final result involves a cancellation.
In particular, the neutralino one-loop contribution to the electron EDM cancels with the stop-associated Barr-Zee contribution for $\mu<0$.
This cancellation leads to the weaker electron EDM constraint in the left panel of \cref{fig:m0mhalf0}, for small $m_0$ and $M_{1/2}$.
This region is instead excluded by the $\mu \to e \gamma$ bounds.
The LFV constraints in this region are enhanced for $\sign(\mu) > 0$, since the Wilson coefficients $C_{LL}$ and $C_{LR}$ in \cref{eq:LFV} have the right signs to add.

\begin{figure}
	\centering
	\includegraphics[width=7.5cm,clip]{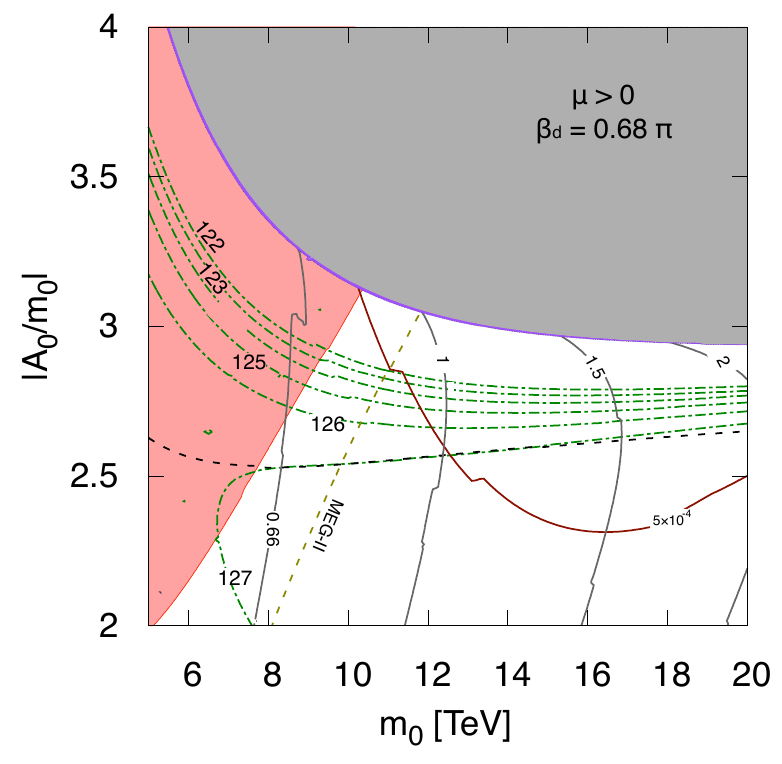}
	\includegraphics[width=7.5cm,clip]{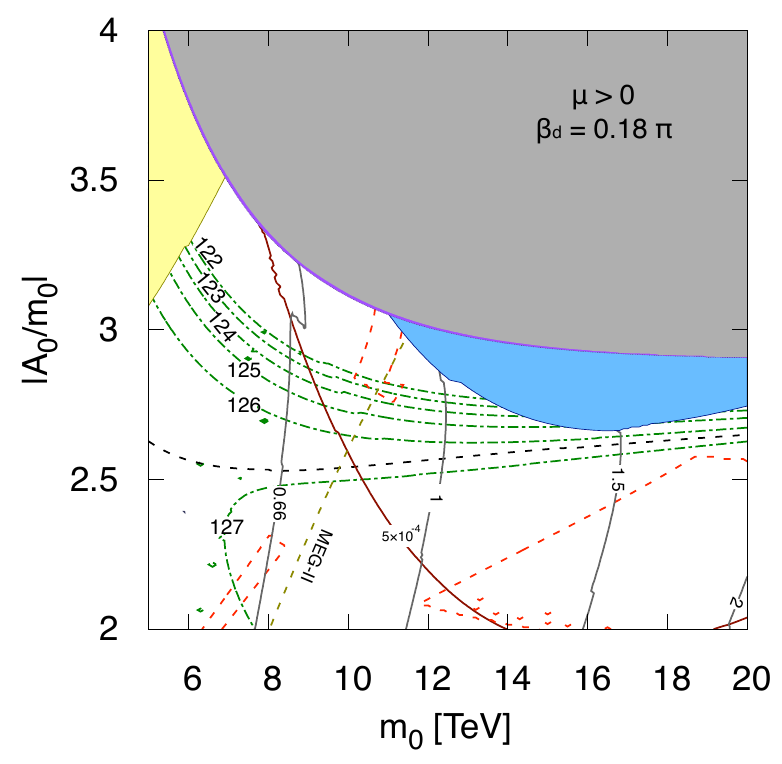} \\
	\includegraphics[width=7.5cm,clip]{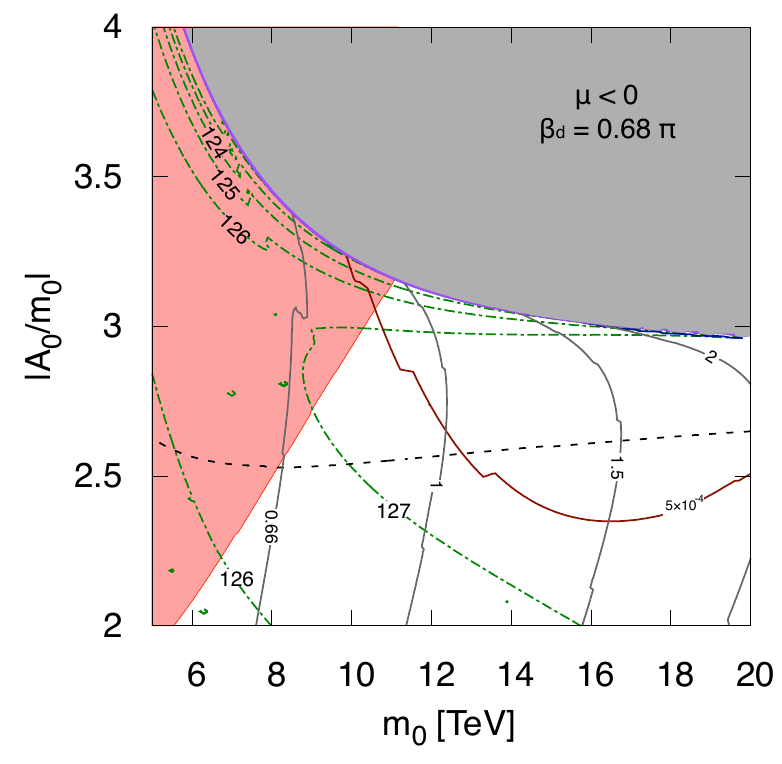}
	\includegraphics[width=7.5cm,clip]{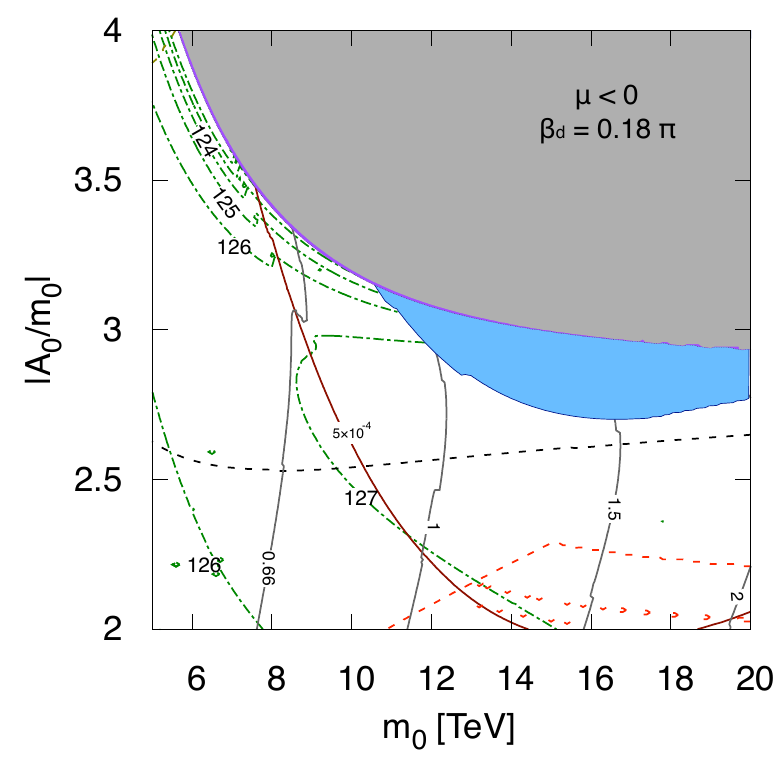}
	\caption{$m_0$-$A_0/m_0$ contour plots.
	$M_{1/2} = 4.5~\mathrm{TeV}$, and $\tan\beta = 6$ are assumed.
	Black-solid lines indicate the partial proton lifetime $\tau(p \to K^+ \ovl \nu)$ in units of $10^{34}$ years.
	Black-shaded regions are excluded by the charge-color breaking (CCB)-charged/colored LSP (CLSP) constraints.
	Above the black-broken lines the $B$-condition is satisfied.
	The green dotted lines indicate the Higgs mass.
	The mass difference $m_{\widetilde t_1} - m_{\chi^0_1}$ is below $100~\text{GeV}$ in the purple shaded region.
	The red and yellow shaded regions are excluded by the current bounds,
	and the red and yellow dashed lines are future limits from the electron EDM and $\mu \to e \gamma$, respectively.
	}\label{fig:m0a01}
\end{figure}

The $B$ matching condition in \cref{eq:Bcond} cannot be satisfied
\footnote{
If the A-terms are complex, this condition may be weakened.
}
for $A_0 = 0$ over the entire plane.
To satisfy the $B$ matching condition, we now consider non-zero $A$-terms.
Non-zero input values for $A_0$ enhance flavor-violation as shown in \cref{eq:offdiag_dsq,eq:offdiag_sl}.

To satisfy the $B$ matching condition, we take the same parameters used in the top panels of \cref{fig:m0mhalf0}, except $A_0 = -3 m_0$, and plot the $m_0$-$M_{1/2}$ plane in the bottom panels of \cref{fig:m0mhalf0}.
The larger $A$-term enhances the LFV in the lepton sector leading to a much larger region excluded by electron EDM constraints, the red region.
The mass difference of the lightest stop and neutralino is below $100~\mathrm{GeV}$ in the purple region, and the gray region with larger $m_0$ is excluded due to a tachyonic stop.
\footnote{
Although we do not calculate it, along this boundary there will be a region where stop-coannihilation can occur.
Since the region excluded by the electron EDM and Kaon oscillation constraints only extends to $M_{1/2}\sim 4$ TeV, bino dark matter from stop coannihilation will extend well beyond the constraints \cite{Ellis:2014ipa,Ellis:2018jyl}.
}
Below the black dashed line in the bottom panels, the $B$ matching condition is satisfied.

In \cref{fig:m0a01}, we show $m_0$-$A_0/m_0$ contour plots with the boundary gaugino mass taken to be $M_{1/2} = 4.5~\mathrm{TeV}$ and $\tan\beta = 6$.
We choose $\beta_d = 0.68 \pi$ , the maximal value for the phase controlling the electron EDM, in the left figure and a more moderate value of $\beta_d = 0.18 \pi$ for the right figure.
We assume positive (negative) $\mu$ in the top (bottom) panels of \cref{fig:m0a01}.
The $B$ matching condition is satisfied above the black dashed lines in each panel.
The gray-shaded regions are excluded because the LSP is colored/charged or it has a charge-color breaking (CCB) minimum due to tachyonic charged sfermion. We will refer to this region as the CCB-CLSP constraints.

The current bound for the electron EDM, $ |d_e| < 9.3 \times 10^{-29}~[e~\text{cm}]$ \cite{Baron:2016obh}, puts the strongest constraint on the parameter space for the figures with $\beta_d = 0.68 \pi$.
In the right-panels, the electron EDM constraints are weaker.
However, the future sensitivity of the ACME experiment can put rather severe constraints on this parameter space.
Most of the parameter space is accessible to the ACME experiments, with the sensitivity indicated by the red-dashed line.
For the left two figures of \cref{fig:m0a01}, the entire parameter space is in reach of the ACME experiments.
The right two figures have some blind spots.
Due to cancellations among the one-loop and two-loop contributions to the electron EDMs for $\mu > 0$, the prediction is beyond the future sensitivity of ACME for the islands around $m_0 \simeq 6\,\text{-}\,10~\mathrm{TeV}$.

Although the EDM constraints become weaker in the figures with $\beta_d = 0.18\pi$, the LFV muon decay still constrains the parameters space, if the Higgsino mass parameter $\mu$ is assumed to be positive.
In particular, the expected future sensitivity at the MEG-II experiment, $Br(\mu \to e \gamma) \simeq 6 \times 10^{-14}$ \cite{Baldini:2018nnn}, will be comparable to the current proton decay constraint.

The future sensitivity of Hyper-Kamiokande experiment to proton decay is reported as $\tau(p\to K^+\ovl \nu) \simeq 2.5 \times 10^{34}$~years at 90\% confidence level \cite{Abe:2011ts}.
Thus, the whole parameter space shown in \cref{fig:m0a01} will be tested by future proton decay experiment.

\begin{table}[t]
	\caption{Sparticle and Higgs Mass Spectrum:}
	\centering
	\label{tab:Spectrum}
	\begin{tabular}{c|c}
		\hline
		Input & \\ \hline
		$m_0$ & $15.1~\text{[TeV]}$ \\
		$M_{1/2}$ & $4.5~\text{[TeV]}$ \\
		$A_0/m_0$ & $-3.02$ \\
		$\tan\beta$ & $6$ \\
		$\text{sign}(\mu)$ & $-1$ \\ \hline
	\end{tabular}
	\begin{tabular}{c|c}
		\hline
		Particle & Mass \\
		\hline
		\hline
		$h$ & $125.2~\text{[GeV]}$ \\
		$H,A,H^\pm$ & $23.0~\text{[TeV]}$ \\
		$(\chi^0_1,\chi^0_2,\chi^0_3,\chi^0_4)$ & $(2.56,4.14,19.2,19.2)~\text{[TeV]}$ \\
		$(\chi^\pm_1,\chi^\pm_2)$ & $(4.14,19.2)~\text{[TeV]}$ \\
		$\widetilde g$ & $8.70~\text{[TeV]}$ \\
		$(\widetilde \nu_1, \widetilde \nu_2, \widetilde \nu_3)$ & $(12.6, 12.8, 15.0)~\text{[TeV]}$ \\
		$(\widetilde \tau_1, \widetilde \tau_2)$ & $(10.4, 13.1)~\text{[TeV]}$ \\
		$(\widetilde e_{L1}, \widetilde e_{L2}, \widetilde e_{R1,2})$ & $(14.2,15.4,16.1)~\text{[TeV]}$ \\
		$(\widetilde t_1, \widetilde t_2)$ & $(2.61,10.9)~\text{[TeV]}$ \\
		$(\widetilde b_1, \widetilde b_2)$ & $(10.9, 12.6)~\text{[TeV]}$ \\
		$(\widetilde u_{L}, \widetilde u_{R})$ & $(16.8, 17.3)~\text{[TeV]}$ \\
		$(\widetilde d_{R1}, \widetilde d_{R2}, \widetilde d_{L1,2})$ & $(15.9,17.0,17.3)~\text{[TeV]}$ \\
		\hline
	\end{tabular}
\end{table}

Lastly, we give the mass spectrum for a reference point in \cref{tab:Spectrum} for a point near the CCB-CLSP boundary in the left-bottom panel of \cref{fig:m0a01}.
In the case of minimal $SU(5)$, it is hard to observe squarks with a mass above $10~\mathrm{TeV}$, even at high-energy colliders with $\sqrt{s} = 100~\mathrm{TeV}$\footnote{Gluinos with a mass below $10~\mathrm{TeV}$ can potentially to be discovered at a 100~TeV colliders \cite{Ellis:2015xba}}.
In contrast to models with CMSSM matter content, minimal SU(5) super-GUTs with right-handed neutrinos, and a large neutrino Yukawa matrix, break the degeneracy of the first two generation sfermions.
If this degeneracy is sufficiently broken, the effects of the first and second generation sfermions may be detectable at future experiments.

\section{Conclusion \label{sec:conclusion}}

In this work, we have revisited the minimal supersymmetric $SU(5)$ with three right-handed neutrinos.
Using the best-fit values for the neutrino mixing angles, the Dirac phase, and the SM parameters, we evaluated the low-scale sparticle mass spectrum using CMSSM-like input masses and calculated the relevant flavor and CP violating signatures.

We have focused on the case with right-handed neutrinos masses around $10^{15}~\mathrm{GeV}$.
The large neutrino Yukawa couplings, large mixing angles, and $CP$ phase for this case induce significant flavor-changing and $CP$-violating processes.
Kaon mixing, the electron EDM, and proton decay are the most important constraints on this model.
The other important experimental constraint, the Higgs mass with $m_h \simeq 125~\mathrm{GeV}$, is compatible with the flavor and $CP$-violating constraints if the $A$-terms are large at the boundary scale.
Although the strongest constraints are currently from proton decay, future experiments, such as MEG-II and ACME experiments, will probe some of the parameter space of these model in a complementary way.

Because of the presences of right-handed neutrinos, there are more GUT-scale phases than in minimal $SU(5)$.
Because these phases are not restricted by low-energy physics, they are, in general, free parameters.
Some of these phases, $\vph_{u_i}$, are used to suppress proton decay, while others, $\vph_{d_i}$, are taken to maximize $CP$ violation for the most stringent bounds.
This work focused on the effects of GUT-scale phases on flavor and $CP$-violation.
However, the $CP$-odd observables also depend on a $CP$ phase in the soft-supersymmetry breaking parameters, which we have set to zero.
These additional phases could lead to cancellations in the contributions to $CP$-odd observables.
We also note that we assumed the normal hierarchical structure for the neutrino masses and a diagonal mass matrix for the right-handed neutrinos.
The constraints from flavor and $CP$ observables are expected to change by a factor of two if the inverted hierarchy is assumed.
The flavor and $CP$ constraints can also change if we take a more generic flavor structure for the right-handed neutrino sector.

Concerning the mass spectrum in the presence of the right-handed neutrinos, the previous studies with CMSSM boundary masses have revealed that the left-handed stau can be the NLSP instead of the right-handed stau.
Contrary to what we expected, this is not possible in super-GUT models with right-handed neutrinos due to radiative corrections involving GUT scale particles and the right-handed neutrinos.
These corrections drive the neutrino soft mass negative above the GUT scale.
The radiative corrections below the GUT scale then have the opposite effect.

Although almost all sparticles are too heavy to be discovered at collider experiments, intensity-frontier experiments can potentially discover and/or constrain these models.

\subsection*{Acknowledgement}
We would like to thank K. A. Olive for useful discussion.
The work of K. K. and T. K. was supported by IBS under the project code, IBS-R018-D1.

\bibliography{ref}
\end{document}